\documentclass[10pt,twocolumn]{extarticle}
\usepackage[a4paper,top=19.5mm,bottom=20mm,left=16mm,right=16mm,columnsep=6.3mm,headheight=13pt,headsep=5mm,footskip=10mm]{geometry}
\usepackage{cuted}
\usepackage[T1]{fontenc}
\usepackage[utf8]{inputenc}
\usepackage{newtxtext,newtxmath}
\usepackage[scaled=.93]{helvet}
\usepackage{microtype}
\usepackage[version=4]{mhchem}
\usepackage{amsmath,graphicx,booktabs,multirow,tabularx,siunitx}
\usepackage{bm}
\usepackage[table]{xcolor}
\usepackage{caption,subcaption,titlesec,dblfloatfix,balance,ragged2e,url}
\usepackage{fancyhdr,lastpage}
\usepackage[numbers,sort&compress]{natbib}
\usepackage[hidelinks]{hyperref}
\definecolor{WileyGray}{RGB}{235,235,235}
\titleformat{\section}{\bfseries\fontsize{11.6}{13.4}\selectfont}{\thesection}{.7em}{}
\titlespacing*{\section}{0pt}{12pt}{5pt}
\titleformat{\subsection}{\bfseries\fontsize{10.4}{12.2}\selectfont}{\thesubsection}{.6em}{}
\titlespacing*{\subsection}{0pt}{8pt}{3pt}

\DeclareCaptionLabelSeparator{wileysep}{\hspace{4pt}\textbar\hspace{6pt}}
\DeclareCaptionFont{wileylabel}{\bfseries\sffamily\footnotesize}
\fancypagestyle{plain}{\pagestyle{fancy}}
\makeatletter
\newcommand{\WileyGraphic}[2][]{%
 \IfFileExists{#2}{\includegraphics[#1]{#2}}{%
  \fbox{\begin{minipage}[c][42mm][c]{.90\linewidth}\centering\small
  Missing figure asset: \texttt{\detokenize{#2}}\end{minipage}}}}
\makeatother
\begin{document}
\twocolumn[
\begin{@twocolumnfalse}
{\sffamily\footnotesize\bfseries\rule{2.5pt}{10pt}\hspace{6pt}REVIEW\par}
\vspace{12pt}
{\bfseries\fontsize{18}{21}\selectfont
Point Defects in Two-Dimensional Materials: From Fundamental Coupling to Functional Properties\par}
\vspace{9pt}
{\fontsize{9.8}{12.2}\selectfont
Nguyen Tuan Hung\textsuperscript{1,*}, 
Nguyen Tran Gia Bao\textsuperscript{2}, 
Renhui Liu\textsuperscript{3,4}, 
Ye Zhang\textsuperscript{3,4}, 
Jianqi Huang\textsuperscript{5}, 
Huaihong Guo\textsuperscript{6}, 
Vu Thi Hanh Thu\textsuperscript{2}, 
Teng Yang\textsuperscript{3,4,5,*}, 
and Riichiro Saito\textsuperscript{7,*}\par}
\vspace{7pt}
{\fontsize{8.5}{10.5}\selectfont
\textsuperscript{1}Frontier Research Institute for Interdisciplinary Sciences, Tohoku University, Sendai, Japan\\
\textsuperscript{2}Faculty of Physics and Physics Engineering, University of Science, Vietnam National University Ho Chi Minh City, Ho Chi Minh City, Viet Nam\\
\textsuperscript{3}School of Materials Science and Engineering, University of Science and Technology of China, Shenyang 110016, China\\
\textsuperscript{4}Shenyang National Laboratory for Materials Science, Institute of Metal Research, Chinese Academy of Sciences, Shenyang 110016, China\\
\textsuperscript{5}Institute of Quantum Materials and Devices, Liaoning Academy of Materials, Shenyang 110167, China\\
\textsuperscript{6}Liaoning Provincial Key Laboratory of Novel Micro-Nano Functional Materials, College of Sciences, Liaoning Petrochemical University, Fushun 113001, China\\
\textsuperscript{7}Department of Physics, Tohoku University, Sendai 980-8578, Japan\\[3pt]
\textbf{Correspondence:} 
Nguyen Tuan Hung (\href{mailto:nguyen.tuan.hung.e4@tohoku.ac.jp}{nguyen.tuan.hung.e4@tohoku.ac.jp}); 
Teng Yang (\href{mailto:yanghaiteng@msn.com}{yanghaiteng@msn.com});
Riichiro Saito (\href{mailto:r.saito.sendai@gmail.com}{r.saito.sendai@gmail.com})\par}
\vspace{8pt}
{\fontsize{8.7}{10.7}\selectfont\noindent\textbf{Keywords:} Point defects \textbar{} electron-defect interaction \textbar{} defect-induced Raman \textbar{} first-principles calculations \textbar{} thermoelectricity\par}
\vspace{7pt}
\noindent\colorbox{WileyGray}{\begin{minipage}{\dimexpr\textwidth-2\fboxsep\relax}
\vspace{5pt}\noindent\textbf{ABSTRACT}\\[3pt]
\fontsize{9.2}{11.4}\selectfont
A century after Frenkel introduced the concept of point defects, defects remain central to semiconductor physics and materials functionality. Their effects are particularly pronounced in two-dimensional (2D) materials, where reduced dimensionality and weak dielectric screening enhance defect-induced changes in electronic, optical, vibrational, and transport properties, while also making their theoretical description challenging. In this review, we highlight recent progress in first-principles calculations of point defects in 2D materials, including available computational methods and codes. We discuss how defects influence carrier mobility, electronic cloaking, thermoelectric transport, and Raman spectra, emphasizing their dual role as performance-limiting imperfections and functional design elements. Finally, we outline key challenges and opportunities, including charged-defect scattering, exciton–defect coupling, and emerging artificial-intelligence approaches toward designing defect engineering in 2D semiconductors.
\vspace{5pt}\end{minipage}}
\vspace{11pt}
\end{@twocolumnfalse}
]

\section{Introduction}
Point defects are local deviations from the atomic arrangement of a crystal, yet they can profoundly influence its charge transport, optical response, and lattice dynamics~\cite{squires2026guidelines,lin2016defect,rao2026structural,fu2025ai,gulo2025quenching,li2019quantized}. A century after Frenkel's seminal discussion of vacancy-interstitial pairs in crystals, point defects remain central to both fundamental solid-state physics and the design of materials with tailored functionality~\citep{frenkel1926}.

In two-dimensional (2D) materials, atoms are located at or near a surface or interface, making the material properties particularly sensitive to vacancies, substitutional impurities, antisites, and adsorbates~\cite{lin2016defect,jiang2019defect}. In addition, dielectric screening in 2D systems is strongly nonlocal and depends on the surrounding environment~\cite{thanh2025hbn}. Consequently, the potentials of charged defects can extend over longer length scales than in typical three-dimensional (3D) crystals. Describing these defects therefore requires a theoretical framework that connects their local atomic and electronic structure with the long-range electrostatic response of the material and its environment~\citep{lischner2019multiscale,lu2022first}.

First-principles calculations provide a powerful means of investigating the structures, charge states, and electronic states of defects at the atomic scale~\cite{freysoldt2014first,ito2022wannier}. However, quantitative predictions for charged defects in 2D materials must account for the dielectric environment, long-range interactions, and finite-size effects arising from periodic boundary conditions~\cite{lu2022first,kaasbjerg2020atomistic}. A central challenge is to describe both the local changes in chemical bonding around a defect and the extended dielectric response of the host material. At the first-principles level, point defects can be treated using two complementary strategies. The supercell approach explicitly includes the defect and captures the associated structural relaxation and defect states, but requires large cells to reduce interactions between periodic images. Alternatively, the defect can be treated as a perturbation to the pristine crystal, and electron-defect scattering matrix elements can be evaluated between the pristine Bloch states using the defect-induced change in the potential~\cite{lu2022first,lu2020ab}. This perturbative approach provides an efficient description of defect scattering, while calculations in defect-containing supercells remain necessary to capture localized defect states and defect-modified vibrational modes~\citep{liu2026defectRaman}.

This Review surveys theoretical descriptions of point defects in 2D materials and recent developments in their first-principles treatment. We first introduce the basic concepts of electron-defect interactions and discuss computational approaches. We then consider how defects affect carrier mobility, electron cloaking, thermoelectric transport, and defect-induced Raman spectra in several 2D materials, such as graphene, transition metal dichalcogenides (MoS$_2$, MoSe$_2$, etc.), Janus MoSSe, and SnSe. We also discuss photoluminescence (PL) arising from transitions between donor and acceptor states in hexagonal boron nitride (hBN) and boron-carbon-nitrogen (BCN) materials. In hBN, for example, the peak positions of defect-induced donor-acceptor PL emissions depend on film thickness and the surrounding dielectric environment, illustrating the role of electron-hole interactions and dielectric screening in defect luminescence~\citep{thanh2025hbn}. Together, these examples show how point defects can both limit material performance and provide functional elements for optical and electronic applications.

The remainder of this Review is organized as follows: Section 3 provides a brief introduction to the formation of point defects in 2D materials, as well as definitions of the various types of defects; Section 3 introduces the fundamentals of electron-defect coupling in 2D materials, covering neutral- and charged-defect scattering and connecting model potentials to first-principles descriptions; Section 4 discusses first-principles approaches and their implementation in computational codes, including Perturbo, EPW, and EDI; Section 5 surveys the effects of point defects on functional properties, including carrier mobility, electron cloaking, thermoelectric transport, Raman spectra, and donor-acceptor PL spectra; Section 6 outlines open challenges and future directions, with emphasis on charged-defect scattering, electron-phonon-defect coupling, and exciton-defect coupling; and finally Section 7 concludes the Review.

\section{Formation and engineering of point defects in two-dimensional materials}

\begin{figure}[t]
  \centering
  \includegraphics[width=0.90\linewidth]{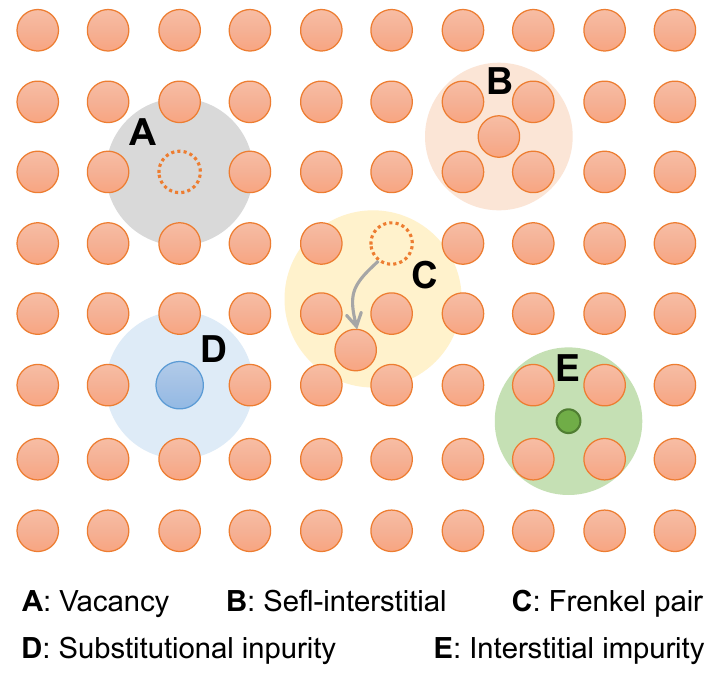}
  \caption{\label{fig:defect} Five common point defects in a simplified two-dimensional crystal lattice, including A: vacancies (an empty lattice site), B: self-interstitial (an extra host atom squeezed between regular sites), C: Frenkel pair (an atom that has left its site for a nearby interstitial position), D: substitutional impurity (a foreign atom on a lattice site, such as a dopant), and E: interstitial impurity (a small foreign atom in the gaps, such as carbon or hydrogen). Local shifts of neighboring atoms are not depicted.}
\end{figure}

Point defects in 2D materials can emerge naturally during synthesis or subsequent processing, and their populations are governed by a combination of thermodynamic stability and kinetic processes~\citet{liang2021defect}. Unlike ideal crystalline models, experimentally synthesized 2D materials inevitably contain vacancies, substitutional impurities, antisites, interstitials, and adsorbates, as shown in Figure~\ref{fig:defect}. Because nearly every atom in an atomically thin material is exposed to the surrounding environment, the formation and evolution of these defects can be particularly sensitive to growth conditions, substrates, chemical environments, and post-synthesis treatments.

\subsection{Defect formation during synthesis}

From a thermodynamic perspective, the probability of forming a particular defect is determined primarily by its formation free energy. In the dilute limit, the equilibrium concentration of a defect species $D$ can be approximately expressed as~\cite{freysoldt2014first}
\begin{equation}
    n_D
    \simeq
    N_{\mathrm{site}}
    \exp
    \left(
    -\frac{\Delta G_f^D}{k_BT}
    \right),
\end{equation}
where $N_{\mathrm{site}}$ is the number of available lattice sites, $\Delta G_f^D$ is the defect formation free energy, $k_B$ is the Boltzmann constant, and $T$ is the growth temperature. Therefore, even relatively small changes in defect formation energy can lead to large variations in defect concentration.

For compound 2D materials, defect formation is also strongly affected by the chemical potentials of the constituent elements. For example, element-rich or element-poor growth conditions can favor different vacancies, antisites, or substitutional defects. Growth temperature, precursor flux, partial pressure, substrate interactions, and growth rate therefore provide important experimental parameters for controlling the resulting defect population. However, the defect structures observed experimentally are not necessarily those predicted solely from thermodynamic equilibrium. The synthesis of 2D materials is often a nonequilibrium process in which defect incorporation, atomic diffusion, migration barriers, and reconstruction at growing edges can determine which defects become kinetically trapped. Rapid cooling after growth may further freeze metastable defect configurations into the material. Therefore, a realistic description of defect formation requires consideration of both formation energetics and kinetic pathways~\cite{komsa2014charged}.

\begin{figure*}[t]
  \centering
  \includegraphics[width=0.75\linewidth]{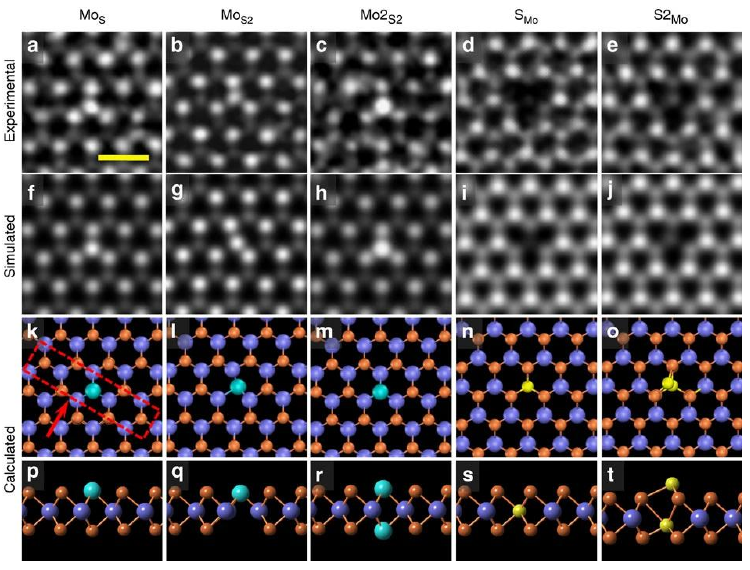}
  \caption{\label{fig:defect-mos2} Several types of the point defect in the MoS$_2$ monolayer. (a–c) High-resolution STEM–ADF images of antisite Mo$_\text{S}$, Mo$_\text{S2}$ and Mo2$_\text{S2}$, respectively. The scale bar is 0.5 nm. (d,e) Atomic structures of antisite defects S$_\text{Mo}$ and S2$_\text{Mo}$, respectively. (f–j) Simulated STEM images based on the theoretically relaxed structures of the corresponding point defects in (a–e). (k–t) Relaxed atomic models of all antisite defects in a–e through DFT calculations, with top and side views, respectively. Light blue is Mo atoms and yellow is S atoms. Reproduced from~\cite{hong2015exploring}. CC BY 4.0.}
\end{figure*}

As shown in Fig.~\ref{fig:defect-mos2}, point defects formed during the synthesis of monolayer MoS$_2$ can adopt several antisite configurations, including Mo$_{\mathrm{S}}$, Mo$_{\mathrm{S_2}}$, Mo$_{2\mathrm{S_2}}$, S$_{\mathrm{Mo}}$, and S$_{2\mathrm{Mo}}$~\cite{hong2015exploring}. Atomic-resolution by combining atomically resolved annular dark-field scanning transmission electron microscopy (ADF-STEM) measurements reveal characteristic contrast patterns for these defects, which are reproduced well by simulated STEM images based on DFT-relaxed atomic structures. In particular, Mo$_{\mathrm{S}}$ and Mo$_{\mathrm{S_2}}$ antisites are found to be dominant in physical vapor deposition (PVD)-grown MoS$_2$, illustrating how synthesis conditions and the associated chemical environment can strongly influence the type and concentration of point defects formed in 2D materials~\cite{hong2015exploring}. The relaxed structures further reveal substantial local atomic reconstruction around some antisites, highlighting the intimate connection between defect formation and local lattice distortion.

Defects may also be introduced or modified after synthesis through thermal annealing, exposure to reactive environments, substrate-induced interactions, oxidation, chemical functionalization, or device fabrication. The experimentally observed defect population should thus be regarded as the outcome of the entire growth and processing history rather than the synthesis step alone.

\subsection{Impact of point defects on electronic properties}

\begin{figure*}[t]
  \centering
  \includegraphics[width=0.95\linewidth]{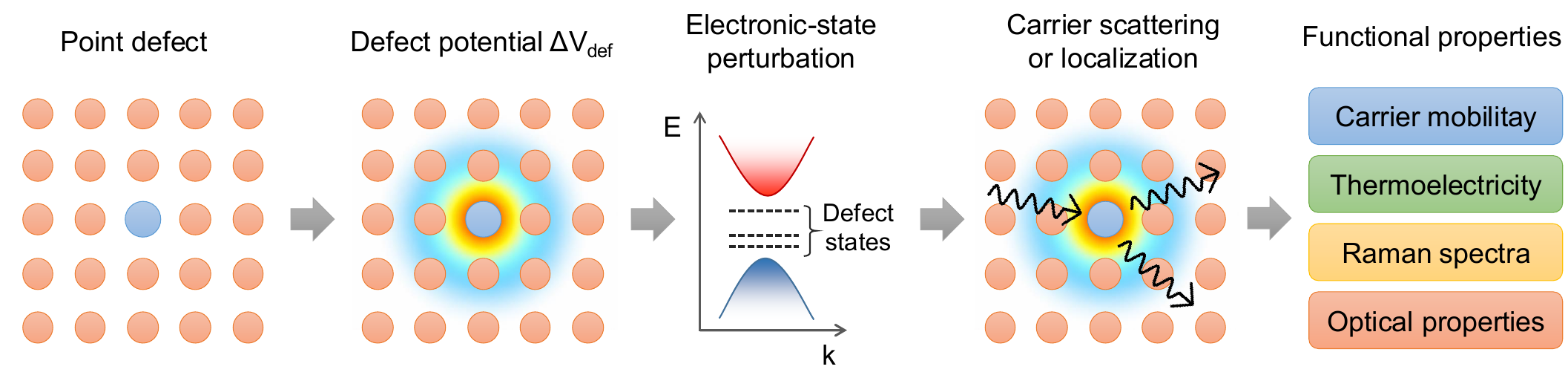}
  \caption{\label{fig:defect-electron} From point defect to functional properties in 2D materials. A point defect (e.g., a substitutional impurity) generates a perturbation of the electronic potential $\Delta V_{\text{def}}$, which modifies the electronic band structure by introducing defect states. The resulting carrier scattering or localization gives rise to changes in functional properties, including carrier mobility, optical response, Raman spectra, and thermoelectric transport.}
\end{figure*}

Once formed, point defects can substantially alter the electronic properties of 2D materials. The removal, addition, or substitution of even a single atom modifies the local bonding environment and can introduce defect-derived electronic states within or near the band gap. Depending on their electronic configuration, these states may behave as donors or acceptors, shift the Fermi level, trap charge carriers, or introduce new optical transitions.

The effect of a defect is determined not only by its atomic structure but also by its charge state. A neutral defect predominantly produces a short-range perturbation associated with changes in local bonding and crystal potential. In contrast, when a defect captures or releases electrons, it becomes charged and generates an additional long-range Coulomb potential. Such charged defects are particularly important in 2D semiconductors because reduced dielectric screening can allow their electrostatic perturbations to extend over comparatively long distances.

Consequently, point defects can influence a wide range of experimentally observable properties. They can limit carrier mobility through electron-defect scattering, modify carrier concentration through unintentional doping, act as nonradiative or radiative recombination centers, introduce localized optical emission, and alter vibrational and thermal-transport properties. At sufficiently high concentrations, defect states may overlap and substantially reconstruct the electronic structure of the host material.

The influence of defects therefore extends beyond the immediate atomic region surrounding the defect. Figure~\ref{fig:defect-electron} illustrates a useful physical picture: a point defect can perturb the electronic potential ($\Delta V_{\text{def}}$), modifying the electronic band structure by introducing defect states. The resulting carrier scattering or localization gives rise to changes in functional properties, including carrier mobility, optical response, Raman spectra, and thermoelectric transport. In Sections 3 and 4, we will explain in more detail the theory and the DFT calculations for defect scattering, and in Section 5, we will show the impact of defect scattering on the functional properties of the 2D materials.

\subsection{From unavoidable defects to defect engineering}

Although point defects have traditionally been regarded as undesirable imperfections that degrade semiconductor performance, advances in synthesis, microscopy, and nanoscale manipulation have increasingly transformed them into controllable degrees of freedom. Defect concentrations can be tuned during growth by controlling stoichiometry and chemical potentials, while post-synthesis techniques such as electron or ion irradiation, plasma treatment, chemical substitution, and thermal annealing can selectively generate, remove, or transform particular defect configurations.

The atomic-scale accessibility of 2D materials provides an especially attractive platform for such defect engineering. Individual defects can in some cases be directly created, imaged, manipulated, or assigned specific charge states. This capability opens opportunities to deliberately use defects to control carrier concentration, scattering, optical emission, magnetism, thermal transport, and other functional properties.

Defect engineering therefore changes the conventional perspective of defects from unavoidable structural imperfections to potentially programmable components of 2D materials. Achieving this goal, however, requires a quantitative understanding of how a particular atomic defect couples to the electronic states of the host material. Such coupling determines whether a defect behaves as a weak perturbation, a carrier-scattering center, a localized electronic state, or a long-range charged impurity. This provides the motivation for examining the fundamental electron-defect coupling mechanisms in 2D materials in the following section.

\section{Fundamentals of electron-defect coupling in two-dimensional materials}
Point defects are unavoidable structural imperfections in 2D materials due to the synthesis process and can strongly perturb their electronic and transport properties. Because all atoms in an atomically thin crystal are located at or near the surface, the electronic states are particularly sensitive to vacancies, substitutional impurities, antisites, and interstitial defects. The reduced dimensionality also modifies both the phase space available for carrier scattering and the dielectric screening of defect potentials compared with three-dimensional (3D) bulk materials.

For an electronic state $\vert n\bm{k}\rangle$ in a pristine 2D crystal, a point defect introduces a perturbation $\Delta V_{\mathrm{def}}$ and couples the initial state $\vert n\bm{k}\rangle$ to a final state $\vert m\bm{k}'\rangle$. The corresponding electron-defect coupling matrix element can be written as
\begin{equation}
\label{eq:2}
\mathcal{M}_{mn}^{\mathrm{def}}(\bm{k}',\bm{k})
=
\left\langle
m\bm{k}'
\left|
\Delta V_{\mathrm{def}}
\right|
n\bm{k}
\right\rangle ,
\end{equation}
where the momentum transfer is
\begin{equation}
\bm{q}=\bm{k}'-\bm{k}.
\end{equation}

Within the Born approximation, the transition probability is determined by Fermi's golden rule as follows~\cite{lundstrom2010fundamentals}:
\begin{equation}
\label{eq:def_general_fgr}
W^{\mathrm{def}}_{n\bm{k}\rightarrow m\bm{k}'}
=
\frac{2\pi}{\hbar}
\left|
\mathcal{M}_{mn}^{\mathrm{def}}(\bm{k}',\bm{k})
\right|^2
\delta
\left(
E_{n\bm{k}}-E_{m\bm{k}'}
\right).
\end{equation}
For transport, scattering processes should additionally be weighted by the momentum-relaxation factor $(1-\cos\theta)$, where $\theta$ is the angle between the initial and final carrier wavevectors. The transport relaxation rate therefore takes the general form~\cite{lundstrom2010fundamentals}
\begin{equation}
\label{eq:def_general_tau}
\frac{1}{\tau_{n\bm{k}}}
=
\sum_{m\bm{k}'}
W^{\mathrm{def}}_{n\bm{k}\rightarrow m\bm{k}'}
(1-\cos\theta).
\end{equation}

Depending on their electronic configuration and charge state, point defects can generate qualitatively different scattering potentials. Neutral defects predominantly produce localized, short-range perturbations, whereas ionized defects additionally generate long-range Coulomb fields. These two limits provide a useful starting point for understanding electron-defect coupling in 2D materials.

\subsection{Neutral-defect scattering}

Neutral vacancies, substitutions, and other localized lattice perturbations can be approximated, in the simplest analytical limit, by a short-range contact potential. For a defect located at the in-plane position $\bm{R}_j$, we write
\begin{equation}
\label{eq:def2d1}
V_j(\bm{\rho})
=
U_0
\delta^{(2)}
(\bm{\rho}-\bm{R}_j),
\end{equation}
where $\bm{\rho}=(x,y)$ is the in-plane coordinate and $U_0$ is the area-integrated strength of the short-range defect potential.

For $N_{\mathrm{def}}$ defects randomly distributed over a 2D crystal with area $A$, the total perturbation is
\begin{equation}
\label{eq:def2d2}
V_{\mathrm{tot}}(\bm{\rho})
=
\sum_{j=1}^{N_{\mathrm{def}}}
U_0
\delta^{(2)}
(\bm{\rho}-\bm{R}_j).
\end{equation}

Using plane-wave states normalized to the 2D area,
\begin{equation}
\langle \bm{\rho}|\bm{k}\rangle
=
\frac{1}{\sqrt{A}}
e^{i\bm{k}\cdot\bm{\rho}},
\end{equation}
the scattering matrix element becomes
\begin{equation}
\label{eq:def2d3}
\mathcal{M}_{\mathrm{def}}^{\mathrm{2D}}
=
\left\langle
\bm{k}'
\left|
V_{\mathrm{tot}}
\right|
\bm{k}
\right\rangle
=
\frac{U_0}{A}
\sum_{j=1}^{N_{\mathrm{def}}}
e^{-i\bm{q}\cdot\bm{R}_j}.
\end{equation}

For uncorrelated defect positions, the interference terms between different defects vanish after configurational averaging. Consequently,
\begin{equation}
\label{eq:def2d4}
\left\langle
\left|
\mathcal{M}_{\mathrm{def}}^{\mathrm{2D}}
\right|^2
\right\rangle_{\mathrm{config}}
=
\frac{N_{\mathrm{def}}U_0^2}{A^2}
=
\frac{n_{\mathrm{def}}^{\mathrm{2D}}U_0^2}{A},
\end{equation}
where
\begin{equation}
n_{\mathrm{def}}^{\mathrm{2D}}
=
\frac{N_{\mathrm{def}}}{A}
\end{equation}
is the defect concentration per unit area.

For an isotropic 2D parabolic band,
\begin{equation}
E(\bm{k})
=
\frac{\hbar^2k^2}{2m^*},
\end{equation}
the momentum-relaxation rate is obtained from Fermi's golden rule as
\begin{equation}
\label{eq:def2d5}
\frac{1}{\tau_{\mathrm{def}}^{\mathrm{2D}}(E)}
=
\frac{2\pi}{\hbar}
n_{\mathrm{def}}^{\mathrm{2D}}
U_0^2
\mathcal{D}_{\mathrm{2D}}(E),
\end{equation}
where $\mathcal{D}_{\mathrm{2D}}(E)$ is the 2D electronic density of states per unit area for the relevant scattering channel.

For a parabolic band,
\begin{equation}
\mathcal{D}_{\mathrm{2D}}(E)
=
\frac{m^*}{2\pi\hbar^2},
\end{equation}
which gives
\begin{equation}
\label{eq:def2d6}
\frac{1}{\tau_{\mathrm{def}}^{\mathrm{2D}}(E)}
=
\frac{
n_{\mathrm{def}}^{\mathrm{2D}}
U_0^2m^*
}{
\hbar^3
}.
\end{equation}

Thus, within the parabolic-band and short-range-potential approximations,
\begin{equation}
\label{eq:def2d7}
\frac{1}{\tau_{\mathrm{def}}^{\mathrm{2D}}(E)}
\propto E^0.
\end{equation}
The absence of an explicit energy dependence originates from the constant electronic density of states of a 2D parabolic band. This behavior differs fundamentally from the $\sqrt{E}$ dependence obtained for the corresponding short-range scattering problem in a 3D parabolic band.

In realistic 2D materials, the defect perturbation is generally more complicated than a delta-function potential. Atomic relaxation, orbital hybridization, intervalley scattering, spin-orbit coupling, and the finite out-of-plane extent of the electronic wavefunctions can introduce significant momentum and band dependence into $\mathcal{M}_{mn}^{\mathrm{def}}(\bm{k}',\bm{k})$. The contact-potential model should therefore be regarded as the simplest analytical limit of the more general first-principles electron-defect matrix element in Eq.~\eqref{eq:2}.

\subsection{Charged-defect scattering}

Point defects can capture or release electrons depending on the Fermi level and their thermodynamic charge-transition levels. An ionized defect with charge
\begin{equation}
q_d=Z_de,
\end{equation}
where $Z_d$ is an integer defect charge state, generates a long-range Coulomb perturbation. The long-range character of this interaction is particularly important in 2D materials because dielectric screening is generally weaker and more spatially nonlocal than in bulk semiconductors.

For a defect located at $z_d$, the in-plane Fourier transform of the Coulomb interaction can be written as~\cite{price1981two,ando1982electronic}
\begin{equation}
\label{eq:def15a}
V_{\mathrm{bare}}^{\mathrm{2D}}
(q,z;z_d)
=
\frac{Z_de^2}
{2\varepsilon_0\kappa_0 q}
e^{-q|z-z_d|},
\end{equation}
where
\begin{equation}
q
=
|\bm{k}'-\bm{k}|
=
2k\sin\left(\frac{\theta}{2}\right)
\end{equation}
is the in-plane momentum transfer, and $\kappa_0$ represents the effective dielectric environment.

The finite out-of-plane extent of the electronic state can be incorporated through a form factor,
\begin{equation}
\label{eq:def17}
\mathcal{F}_{\mathrm{2D}}(q)
=
\iint
|\psi(z)|^2
|\psi(z_d)|^2
e^{-q|z-z_d|}
\,\mathrm{d}z\,\mathrm{d}z_d .
\end{equation}
The corresponding effective bare interaction is therefore
\begin{equation}
\label{eq:def15b}
U_{\mathrm{bare}}^{\mathrm{2D}}(q)
=
\frac{Z_de^2}
{2\varepsilon_0\kappa_0q}
\mathcal{F}_{\mathrm{2D}}(q).
\end{equation}

For an ideal 2D electron gas, static screening can be introduced through the Stern-Howard dielectric function~\cite{stern1967properties},
\begin{equation}
\label{eq:def_dielectric_2d}
\epsilon_{\mathrm{2D}}(q)
=
1+
\frac{q_s^{\mathrm{2D}}}{q},
\end{equation}
where the 2D screening wavevector is
\begin{equation}
\label{eq:q2D}
q_s^{\mathrm{2D}}
=
\frac{e^2}
{2\varepsilon_0\kappa_0}
\mathcal{D}_{\mathrm{2D}}(E_F).
\end{equation}

The screened charged-defect interaction is then
\begin{equation}
\label{eq:def16}
U_{\mathrm{def}}^{\mathrm{2D}}(q)
=
\frac{
Z_de^2
}{
2\varepsilon_0\kappa_0
\left(q+q_s^{\mathrm{2D}}\right)
}
\mathcal{F}_{\mathrm{2D}}(q).
\end{equation}

For a degenerate 2D electron gas with a parabolic dispersion, $\mathcal{D}_{\mathrm{2D}}(E_F)$ is independent of energy. Consequently, $q_s^{\mathrm{2D}}$ is independent of carrier density within this idealized zero-temperature Thomas-Fermi/Stern limit. In realistic 2D semiconductors, however, finite temperature, multivalley electronic structure, nonparabolic bands, dielectric anisotropy, and nonlocal screening can substantially modify this simple behavior.

For randomly distributed charged defects with areal concentration $n_{\mathrm{def}}^{\mathrm{2D}}$, the momentum-relaxation rate is
\begin{equation}
\label{eq:def18}
\begin{split}
\frac{1}
{\tau_{\mathrm{def}}^{\mathrm{2D}}(E)}
=
\frac{2\pi}{\hbar}
n_{\mathrm{def}}^{\mathrm{2D}}
\int &
\frac{\mathrm{d}^2\bm{k}'}
{(2\pi)^2}
\left|
U_{\mathrm{def}}^{\mathrm{2D}}(q)
\right|^2 \\
&\times
(1-\cos\theta)
\delta
\left[
E(\bm{k})-E(\bm{k}')
\right].
\end{split}
\end{equation}

For the parabolic dispersion
$E=\hbar^2k^2/(2m^*)$, integration over the magnitude of $\bm{k}'$ gives
\begin{equation}
\label{eq:def19}
\frac{1}
{\tau_{\mathrm{def}}^{\mathrm{2D}}(E)}
=
\frac{
m^*n_{\mathrm{def}}^{\mathrm{2D}}
}
{2\pi\hbar^3}
\int_0^{2\pi}
\left|
U_{\mathrm{def}}^{\mathrm{2D}}(q)
\right|^2
(1-\cos\theta)
\,\mathrm{d}\theta .
\end{equation}

Substituting Eq.~\eqref{eq:def16} gives
\begin{equation}
\label{eq:def20}
\frac{1}
{\tau_{\mathrm{def}}^{\mathrm{2D}}(E)}
=
\frac{
m^*
n_{\mathrm{def}}^{\mathrm{2D}}
Z_d^2e^4
}
{
8\pi
\hbar^3
\varepsilon_0^2
\kappa_0^2
}
\int_0^{2\pi}
\frac{
(1-\cos\theta)
\left|
\mathcal{F}_{\mathrm{2D}}(q)
\right|^2
}
{
\left[
2k\sin(\theta/2)
+
q_s^{\mathrm{2D}}
\right]^2
}
\,\mathrm{d}\theta .
\end{equation}

Equation~\eqref{eq:def20} highlights several characteristic features of charged-defect scattering in 2D systems. First, the bare Coulomb interaction strongly favors small-momentum-transfer scattering. Second, dielectric screening suppresses this divergence through $q_s^{\mathrm{2D}}$. Third, the transport factor $(1-\cos\theta)$ reduces the contribution from small-angle forward scattering because such events only weakly relax the carrier momentum.

For atomically thin semiconductors, the screening physics can be substantially more complex than the simple Stern-Howard model. The dielectric response may depend strongly on wavevector, layer thickness, substrate environment, and the intrinsic polarizability of the 2D crystal. Consequently, quantitative calculations of charged-defect scattering generally require the full momentum-dependent screened defect potential rather than a single macroscopic dielectric constant.

\subsection{From model potentials to first-principles electron-defect coupling}

The analytical expressions above provide useful physical insight into how dimensionality, defect charge, screening, and electronic phase space control electron-defect scattering. However, realistic point defects in 2D materials are neither ideal delta-function scatterers nor simple point charges. A defect can simultaneously produce local changes in bonding, lattice relaxation, charge redistribution, electrostatic fields, and orbital hybridization.

In first-principles calculations based on DFT, these effects are naturally contained in the defect-induced perturbation potential,
\begin{equation}
\label{eq:def_deltaV}
\Delta V_{\mathrm{def}}
=
V_{\mathrm{KS}}^{\mathrm{def}}
-
V_{\mathrm{KS}}^{\mathrm{pris}},
\end{equation}
where $V_{\mathrm{KS}}^{\mathrm{def}}$ and
$V_{\mathrm{KS}}^{\mathrm{pris}}$ are the Kohn-Sham potentials of the defective and pristine systems, respectively. The microscopic electron-defect (e-d) matrix elements can then be evaluated by Eq.~\eqref{eq:2}. The DFT allows intravalley and intervalley processes, anisotropic electronic structures, multiple bands, spin-orbit coupling, and the microscopic dielectric response to be treated explicitly. The accurate evaluation of these matrix elements, particularly for charged defects and long-range electrostatic perturbations, nevertheless remains one of the central computational challenges in first-principles defect physics of 2D materials. These challenges come from both computational cost and the lack of a theoretical formulation amenable to computation. Computational cost is a key barrier for defect calculations in DFT. As shown in Eq.~\eqref{eq:def_deltaV}, the defect must be placed in a supercell large enough to avoid spurious interactions with image cells due to periodic boundary conditions. As a result, defect calculations may require simulation cells with hundreds of atoms treated explicitly, making them computationally demanding in many DFT codes. In Sec. 4, we will present techniques within DFT that can address these challenges.

\section{Electron-defect coupling from first-principles calculations}
In Sec. 3.3, we introduced the defect-induced perturbation potential for point defects from first-principles calculations. In this section, we discuss in detail the technical methodology for implementing electron-defect calculations and focus primarily on charge-neutral point defects within existing codes, including Perturbo and EDI. 

\subsection{Direct evaluation of electron-defect coupling from DFT}
By using norm-conserving pseudopotentials in the Kleinman-Bylander form~\cite{martin2020electronic}, the Kohn-Sham potentials in Eq.~\eqref{eq:def_deltaV} can be written as a sum of local and nonlocal parts~\cite{kleinman1982efficacious}:
\begin{equation}
\label{eq:29}
V_{\mathrm{KS}}=V_{\mathrm{L}}(\bm r)+\hat{V}_{\mathrm{NL}},
\end{equation}
where the local potential $V_{\mathrm{L}}(\bm r)=V_H(\bm r)+V_{\mathrm{XC}}(\bm r) +V_{\mathrm{pp}}(\bm r)$ contains the Hartree $V_H(\bm r)$, exchange-correlation $V_{\mathrm{XC}}(\bm r)$, and local pseudopotential contributions $V_{\mathrm{pp}}(\bm r)$, while $\hat V_{\mathrm{NL}}$ contains the nonlocal pseudopotential projectors.
Accordingly, the e-d matrix element can also be decomposed into a local and a nonlocal part as 
\begin{equation}
\mathcal{M}_{mn}^{\mathrm{def}}(\bm{k}',\bm{k}) = \mathcal{M}_{mn}^{\mathrm{L}}(\bm{k}',\bm{k})
+\mathcal{M}_{mn}^{\mathrm{NL}}(\bm{k}',\bm{k}),
\label{eq:30}
\end{equation}
with
\begin{equation}
\mathcal{M}_{mn}^{\mathrm{L}}(\bm{k}',\bm{k})=\left\langle m\bm{k}'\left|\Delta V_\mathrm{L}\right|n\bm{k}\right\rangle
\label{eq:31}
\end{equation}
and
\begin{equation}
\mathcal{M}_{mn}^{\mathrm{NL}}(\bm{k}',\bm{k})=\left\langle m\bm{k}'\left|\Delta \hat V_{\mathrm{NL}}\right|n\bm{k}\right\rangle .
\label{eq:32}
\end{equation}
Here, $\Delta V_{\mathrm{L}}(\bm r)=V_{\mathrm{L}}^{\mathrm{def}}-V_{\mathrm{L}}^{\mathrm{pris}}$ and $\Delta \hat{V}_{\mathrm{NL}}(\bm r)=\hat{V}_{\mathrm{NL}}^{\mathrm{def}}-\hat{V}_{\mathrm{NL}}^{\mathrm{pris}}$.
In the all-supercell method, both the defect perturbation potential and the unperturbed Bloch states are obtained from the supercell, which is computationally expensive. 

\begin{figure}[t]
    \centering
    \includegraphics[width=1.0\linewidth]{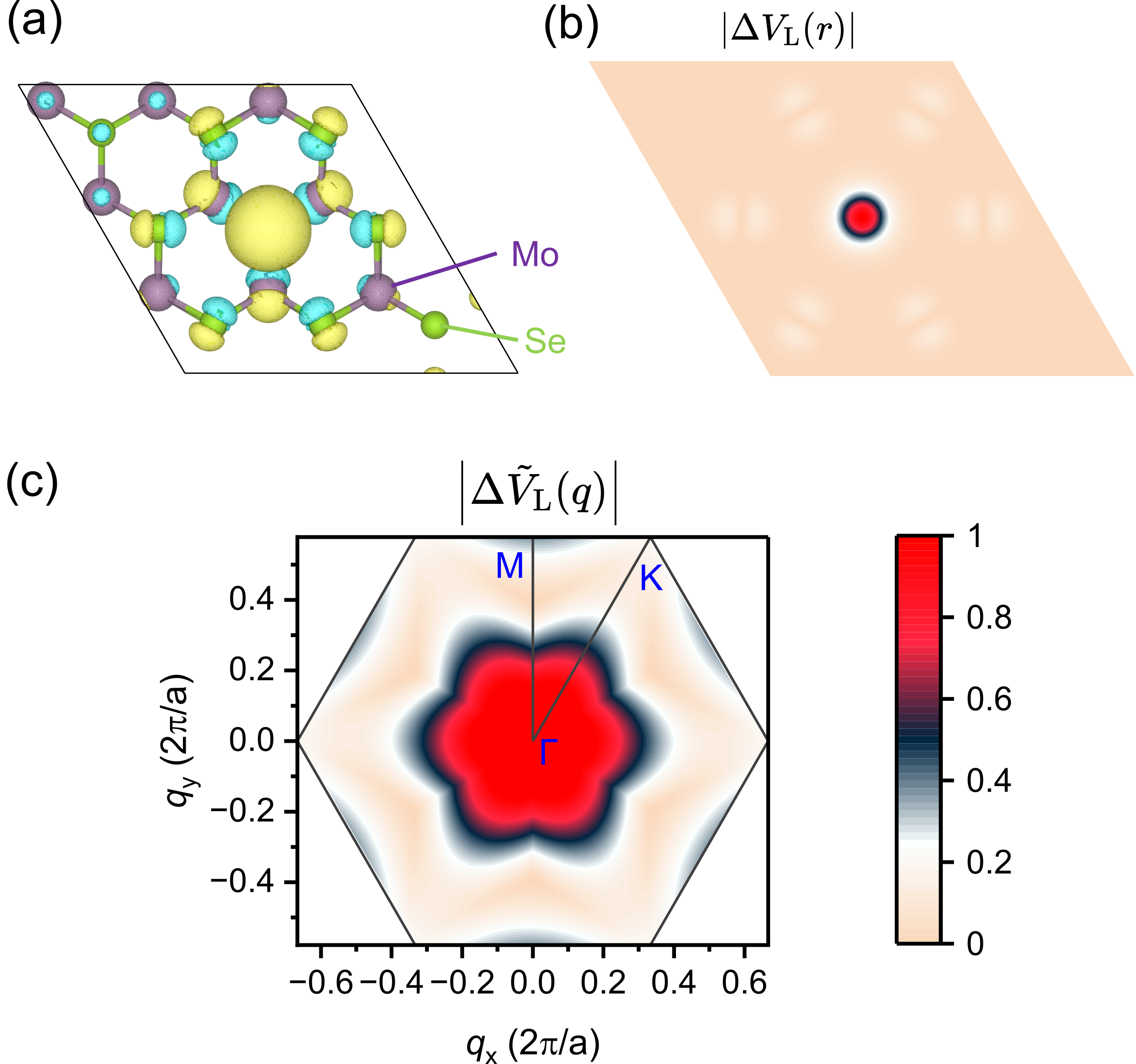}
    \caption{(a) Top-view of local perturbation potential $\Delta V_\mathrm{L}(\bm{r})$ isosurface ($\left|\Delta V_\mathrm{L}(\bm{r})\right|=\SI{0.60}{eV}$) in the $3 \times 3$ MoSe$_2$ monolayer supercell containing a single Se vacancy. Yellow and green surfaces denote positive and negative values of $\Delta V_\mathrm{L}(\bm{r})$, respectively. (b) Cutting-plane view of normalized $\left|\Delta V_\mathrm{L}(\bm{r})\right|$ passing through the Se vacancy site.  (c) Normalized magnitude of the Fourier coefficients of the local perturbation potential, $\left|\Delta \widetilde{V}_\mathrm{L}(\bm{q})\right|$, in the unit-cell Brillouin zone (BZ). The values shown in (b) and (c) are normalized by their corresponding maximum magnitudes and share the same color bar. Reproduced from Ref.~\citep{liu2026defectRaman} with permission. Copyright 2026 American Physical Society.}
    \label{fig:4}
\end{figure}

A computationally efficient strategy is to obtain the defect perturbation from the supercell while evaluating the Bloch states in the primitive-unit-cell~\cite {lu2019efficient}. For the local component, the defect potential can be Fourier transformed as
\begin{equation}
\Delta \widetilde{V}_\mathrm{L}(\bm q)=
\frac{1}{\Omega_{\mathrm{uc}}}
\int_{\Omega_{\mathrm{sup}}}
\Delta V_L(\bm{r})
e^{-i\bm{q}\cdot\bm{r}}\mathrm{d}\bm r,
\label{eq:33}
\end{equation}
where $\Omega_{\mathrm{uc}}$ and $\Omega_{\mathrm{sup}}$ denote the primitive-unit-cell and supercell volumes, respectively. The local e-d matrix element can then be expressed in terms of primitive-unit-cell wave functions as
\begin{equation}
\mathcal{M}_{mn}^{\mathrm{L}}(\bm{k}',\bm{k})=\sum_{\bm{G}}
\Delta\widetilde V_L
(\bm{k}'-\bm{k}-\bm{G})
\left\langle
u_{m\bm{k}'}
\left|
e^{-i\bm{G}\cdot\bm{r}}
\right|
u_{n\bm{k}}
\right\rangle_{\mathrm{uc}},
\label{eq:direct_local_matrix}
\end{equation}
where $\bm q = \bm{k}'-\bm{k}-\bm{G}$ is the transferred momentum in the $|n\bm k\rangle \to |m \bm{k}'\rangle$ scattering process. $u_{n\bm{k}}$ is the unit-cell-periodic part of the Bloch state and $\bm{G}$ is a reciprocal-lattice vector. We thus separate the calculation naturally into two components: the Fourier coefficient of the defect perturbation $\Delta V_L(\bm{r})$ in Eq.~\eqref{eq:33}, which is obtained from the defective- and pristine-supercells, and the electronic matrix elements in Eq.~\eqref{eq:direct_local_matrix}, which are evaluated using primitive-unit-cell Bloch states. For a neutral point defect, $\Delta V_L(\bm{r})$ is localized in real space, and its Fourier components generally decay rapidly with increasing momentum (see Fig.~\ref{fig:4}(b)). Consequently, only a limited number of reciprocal-lattice vectors may be required for convergence.

In Fig.~\ref{fig:4}(a), we present the top view of local perturbation potential $\Delta V_\mathrm{L}(\bm{r})$ isosurface ($\left|\Delta V_\mathrm{L}(\bm{r})\right|=\SI{0.60}{eV}$) in in the $3 \times 3$ MoSe$_2$ monolayer supercell containing a Se vacancy. Yellow and green surfaces denote positive and negative values of $\Delta V_\mathrm{L}(\bm{r})$, respectively. Figure~\ref{fig:4}(b) shows the cutting-plane view of normalized $\left|\Delta V_\mathrm{L}(\bm{r})\right|$ along the $z$ direction passing through the vacancy site, which shows that $\Delta V_\mathrm{L}(\bm{r})$ is strongly localized around the vacancy and decays rapidly away from the defect. In Fig.~\ref{fig:4}(c), we present the magnitude of the Fourier coefficients of the local perturbation potential $\left|\Delta \widetilde{V}_\mathrm{L}(\bm{q})\right|$ in the unit-cell BZ, as defined in Eq.~\eqref{eq:33}. The values are normalized to the maximum magnitude. The Fourier coefficients are mainly concentrated around the $\Gamma$ and M points, with only weak contributions from elsewhere in the BZ.

The nonlocal matrix elements $\mathcal{M}_{mn}^{\mathrm{NL}}(\bm{k}',\bm{k})$ in Eq.~\eqref{eq:32} are calculated by the difference between the nonlocal potentials of a defective-supercell and a pristine-supercell as
\begin{equation}
\mathcal{M}_{mn}^{\mathrm{NL}}(\bm{k}',\bm{k})=\left\langle m\bm{k}'\left| \hat{V}_{\mathrm{NL}}^{\mathrm{def}}\right|n\bm{k}\right\rangle -\left\langle m\bm{k}'\left| \hat{V}_{\mathrm{NL}}^{\mathrm{pris}}\right|n\bm{k}\right\rangle ,
\label{eq:35}
\end{equation}
where the nonlocal potential $\hat{V}_{\mathrm{NL}}^{\alpha}$ with $(\alpha = \mathrm{def}$ or pris) is defined as a sum over all atoms in the supercell of the Kleinman–Bylander projectors $\left|\beta_i^{s,\alpha}\right\rangle$, which are localized in the core region of the $s$-th atom in the supercell:
\begin{equation}
\hat{V}_{\mathrm{NL}}^{\alpha}=\sum_{s,i,j}D_{ij}^{s,\alpha}\left|\beta_i^{s,\alpha}\right\rangle \left\langle\beta_i^{s,\alpha}\right|,
\label{eq:36}
\end{equation}
where $i$ and $j$ are orbital angular momentum quantum numbers, and $D_{ij}^s$ is the Kleinman–Bylander coefficients~\cite{kleinman1982efficacious}. By substituting Eq.~\eqref{eq:36} into Eq.~\eqref{eq:35}, the matrix elements of $\hat{V}_{\mathrm{NL}}^{\alpha}$ can be expressed as
\begin{equation}
\left\langle m\bm{k}'\left| \hat{V}_{\mathrm{NL}}^{\alpha}\right|n\bm{k}\right\rangle=\sum_{s,i,j}D_{ij}^{s,\alpha}\left\langle m\bm{k}'\left|\beta_i^{s,\alpha}\right\rangle \left\langle\beta_j^{s,\alpha}\right| n\bm{k}\right\rangle.
\label{eq:37}
\end{equation}

Similar to the local matrix elements, Eq.~\eqref{eq:37} can split into a
primitive-unit-cell and a supercell calculation by expressing the scalar products as
\begin{equation}
\left\langle\beta_j^{s,\alpha}| n\bm{k}\right\rangle = \frac{1}{\sqrt{\Omega_{\mathrm{uc}}}}\sum_{\bm G}{B^{s,\alpha}_{j\bm k}}^*(\bm G)\left\langle e^{i\bm{G}\cdot\bm{k}}| u_{n\bm{k}}\right\rangle_{\mathrm{uc}},
\label{eq:38}
\end{equation}
where $B^{s,\alpha}_{j\bm k}$ is the Fourier coefficient of the Kleinman–Bylander projector $\beta_j^{s}$ related to the primitive-unit-cell reciprocal lattice vector $\bm G$ as:
\begin{equation}
B^{s,\alpha}_{j\bm k}(\bm G)=\frac{1}{\sqrt{\Omega_{\mathrm{uc}}}}\int_{\Omega_{\mathrm{sup}}}\left[\beta_j^{s,\alpha}(\bm r)e^{-i\bm{k}\cdot\bm{r}}\right]e^{-i\bm{G}\cdot\bm{r}}\mathrm{d}\bm{r}.
\label{eq:39}
\end{equation}

We note that, in general, the nonlocal matrix elements in Eq.~\eqref{eq:35} are nonzero since the defective supercell differs from the pristine one through atomic relaxation and, for vacancies or impurities, through changes in the number or type of atoms. After structural optimization, the relaxed atomic positions and chemical species in the defective supercell determine the spatial locations and species dependence of the corresponding  Kleinman–Bylander projector $\beta_j^{s}$. The radial forms of the $\beta_j^{s}$ and their coefficients $D_{ij}^s$ are provided by the norm-conserving pseudopotentials associated with each atomic species. Consequently, atomic relaxation modifies the phase and spatial dependence of the site-centered projectors, whereas a vacancy or substitutional impurity may additionally remove or replace the set of $\beta_j^{s}$ and $D_{ij}^s$ coefficients associated with the corresponding atomic site. These differences between the defective and pristine supercells give rise to a finite nonlocal contribution $\mathcal{M}_{mn}^{\mathrm{NL}}(\bm{k}',\bm{k})$ to the electron-defect interaction.

Once the matrix elements $\mathcal{M}_{mn}^{\mathrm{def}}(\bm{k}',\bm{k})$ are obtained, the elastic e-d scattering rate can be evaluated within Fermi's golden rule as:
\begin{equation}
\frac{1}{\tau_{n\bm{k}}}=
\frac{2\pi}{\hbar}\frac{n_{\text{at}}C_{\mathrm{d}}}{N_{\bm{k}'}}\sum_{m\bm{k}'}|\mathcal{M}_{mn}^{\mathrm{def}}(\bm{k}',\bm{k})|^2\delta(E_{n\bm{k}}-E_{m\bm{k}'}),
\label{eq:40}
\end{equation}
where $C_\mathrm{d}$ is the defect atomic concentration, $n_{\mathrm{at}}$ is the number of atoms in the primitive-unit-cell, and $N_{\bm{k}'}$ is the number of final-state $\bm{k}'$ points. Equation~(\ref{eq:40}) establishes a direct connection between the defect-induced microscopic perturbation and the carrier relaxation time.

\subsection{Wannier-interpolated electron-defect coupling}
Although the e-d matrix elements can be evaluated directly using the DFT formulation in Eq.~\eqref{eq:30}, calculations of relaxation times in Eq.~\eqref{eq:40} and defect-limited transport require summations over large numbers of pairs $(\bm{k},\bm{k}')$. Direct evaluation on extremely dense Brillouin-zone grids therefore rapidly becomes computationally prohibitive. This limitation can be overcome by combining first-principles DFT calculations with maximally localized Wannier functions (MLWFs), as implemented, for example, in the Wannier90 package~\cite{mostofi2008wannier90,pizzi2020wannier90}.

The central idea is that DFT is first used to compute the e-d matrix elements on a relatively coarse Brillouin-zone grid. These matrix elements are subsequently transformed from the extended Bloch representation into a localized Wannier representation. Because both the Wannier functions and the perturbation generated by a neutral point defect are localized in real space, the resulting e-d matrix elements decay rapidly with distance. This locality makes it possible to reconstruct the interaction efficiently at arbitrary points on a much finer $\bm{k}$ grid.

The pristine Bloch states can be expressed in terms of Wannier functions as
\begin{equation}
|n\bm{k}\rangle=
\sum_{j\bm{R}}
e^{i\bm{k}\cdot\bm{R}}
U_{jn,\bm{k}}^{\dagger}
|j\bm{R}\rangle ,
\label{eq:41}
\end{equation}
where $|j\bm{R}\rangle$ denotes the $j$-th Wannier function centered in the unit cell at lattice vector $\bm{R}$ and $U_{\bm{k}}$ is the unitary transformation connecting the Bloch and Wannier representations. The corresponding e-d matrix elements in real space are
\begin{equation}
\mathcal{M}^{\mathrm{def}}_{ij}(\bm{R}',\bm{R})=
\left\langle
i\bm{R}'
\left|
\Delta V_{\mathrm{def}}
\right|
j\bm{R}
\right\rangle .
\label{eq:42}
\end{equation}

Starting from e-d matrix elements, $\mathcal{M}_{mn}^{\mathrm{def}}(\bm{k}'_c,\bm{k}_c)$, computed directly by DFT on a coarse grid $\bm{k}_c$, the interaction is transformed into the Wannier representation through a generalized double Fourier transform~\cite{lu2020ab},
\begin{equation}
\begin{split}
\mathcal{M}^{\mathrm{def}}_{ij}(\bm{R}',\bm{R})=&
\frac{1}{N_{\bm{k}_c}^{2}}
\sum_{\bm{k}'_c,\bm{k}_c}
e^{i(\bm{k}'_c\cdot\bm{R}'
-\bm{k}_c\cdot\bm{R})} \\
&\times 
\left[U_{\bm{k}'_c}^{\dagger}
\mathcal{M}_{mn}^{\mathrm{def}}(\bm{k}'_c,\bm{k}_c)
U_{\bm{k}_c}\right]_{ij}
\end{split}.
\label{eq:43}
\end{equation}
For a short-range neutral-defect perturbation, $\mathcal{M}_{mn}^{\mathrm{def}}(\bm{R},\bm{R}')$ decreases rapidly as the Wannier centers move away from the defect. Consequently, only matrix elements within a finite real-space region need to be retained. The localized representation therefore acts as a compact description of the full electron-defect interaction.

The electronic Hamiltonian is independently interpolated in the Wannier basis,
\begin{equation}
H(\bm{k}_f)=
\sum_{\bm{R}}
e^{i\bm{k}_f\cdot\bm{R}}
H(\bm{R}),
\label{eq:44}
\end{equation}
and diagonalization of $H(\bm{k}_f)$ provides both the band energies and unitary matrices $U_{\bm{k}_f}$ at arbitrary fine-grid points $\bm{k}_f$. The e-d matrix elements can then be transformed back into the Bloch representation on the fine grid according to~\cite{lu2020ab}
\begin{equation}
\begin{split}
\mathcal{M}_{mn}^{\mathrm{def}}(\bm{k}'_f,\bm{k}_f)=&
\sum_{\bm{R}',\bm{R}}
e^{-i(\bm{k}'_f\cdot\bm{R}'
-\bm{k}_f\cdot\bm{R})}\\
& \times
\left[U_{\bm{k}'_f}
\mathcal{M}_{ij}^{\mathrm{def}}(\bm{R}',\bm{R})
U_{\bm{k}_f}^{\dagger}\right]_{mn}
\end{split}.
\label{eq:45}
\end{equation}

\begin{figure}[t]
    \centering
    \includegraphics[width=1.0\linewidth]{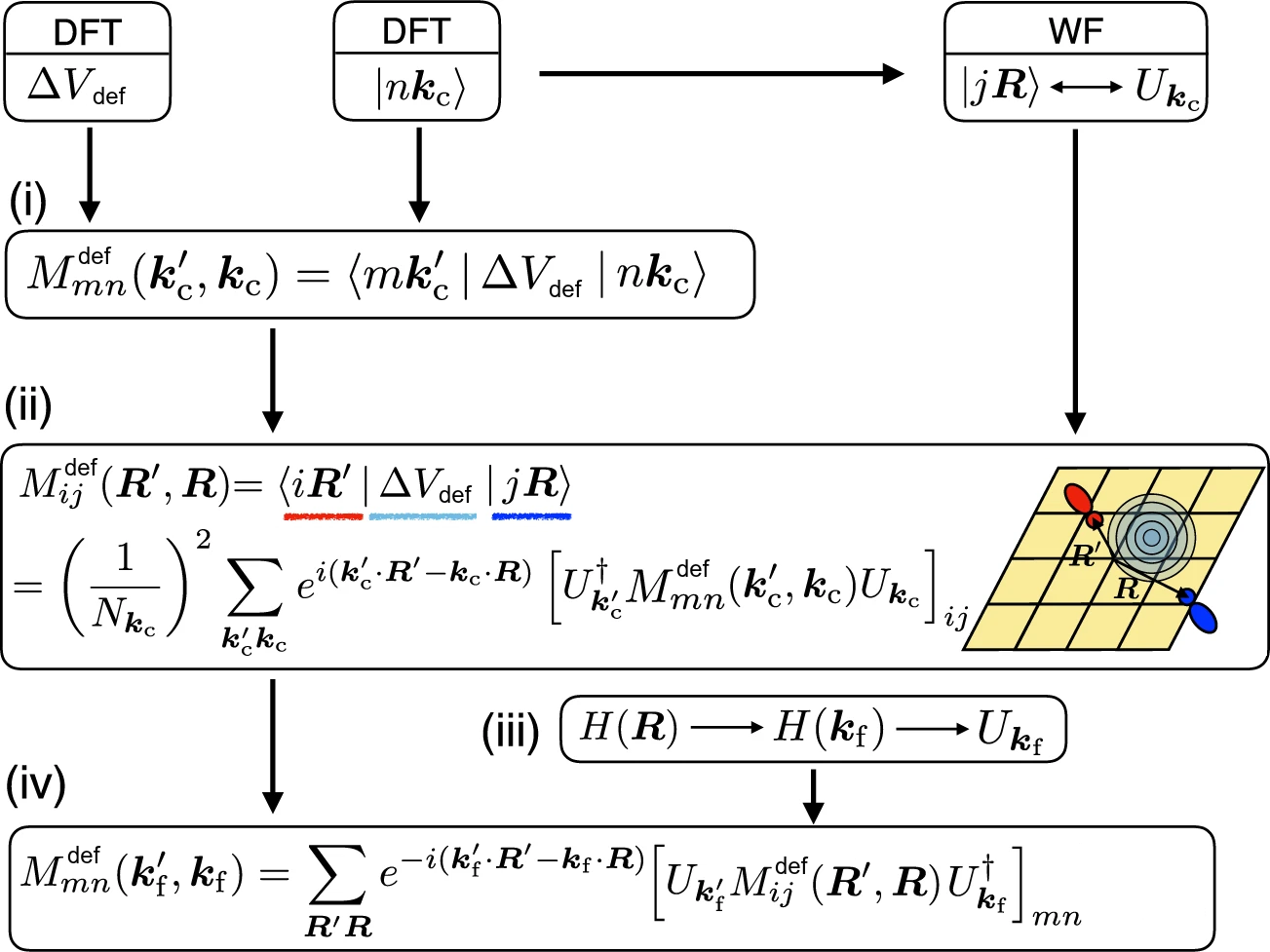}
    \caption{Workflow for interpolating the e–d matrix elements using Wannier interpolation in the Perturbo code~\cite{zhou2021perturbo}. Reproduced from Ref.~\citep{lu2020ab} with permission. CC BY 4.0.}
    \label{fig:5}
\end{figure}

Figure~\ref{fig:5} shows the workflow for interpolating the e-d matrix elements in the Perturbo code~\cite{zhou2021perturbo} using Wannier interpolation, which includes several steps: (i) compute the e-d matrix elements in the Bloch representation on a coarse BZ grid using Eq.~\eqref{eq:30}; (ii) obtain the e-d matrix elements in the Wannier representation using Eq.~\eqref{eq:43}; (iii) interpolate the Hamiltonian using Eq.~\eqref{eq:44} and diagonalize it to obtain the fine-grid unitary matrices $U_{\bm{k}_f}$; (iv) interpolate the e-d matrix elements to any desired pair of fine-grid points using Eq.~\eqref{eq:45}.

Importantly, the Wannier procedure does not introduce a different physical model for the electron-defect interaction. The defect perturbation and the matrix elements on the coarse grid remain first-principles quantities obtained from DFT. Wannier90 provides a localized representation that allows these quantities, together with the electronic band energies and velocities, to be interpolated efficiently to arbitrary $\bm{k}$ points. Therefore, the accuracy of this approach relies on the localization of both the Wannier functions and the defect perturbation. For neutral point defects, the matrix elements in the Wannier representation typically decay over only a few unit cells, making the interpolation particularly efficient. Convergence can then be systematically controlled through the coarse $\bm{k}$ grid used for Wannierization and the spatial extent of the retained real-space matrix elements. 

This approach is especially advantageous for defect-limited transport calculations because the scattering rate in Eq.~(\ref{eq:40}) requires integration over a large number of possible final states. Wannier interpolation makes it possible to evaluate the electron-defect matrix elements on extremely dense uniform or random Brillouin-zone grids at a fraction of the cost of their direct evaluation~\cite{lu2020ab}. The resulting relaxation times can subsequently be combined with Wannier-interpolated band velocities to evaluate the transport distribution function and quantities such as conductivity, mobility, and resistivity.

Thus, the direct DFT and Wannier-interpolated approaches should be regarded as complementary rather than distinct physical theories. Direct DFT provides the fundamental defect perturbation and electron-defect coupling, whereas the Wannier representation supplies an efficient bridge between accurate first-principles calculations on coarse grids and the dense Brillouin-zone sampling required for quantitative carrier-scattering and transport calculations.

\subsection{Computational codes for electron–defect interactions}

Several computational packages have been developed to describe electron-defect or impurity scattering and its influence on carrier transport. Table~\ref{tab:ed_codes} summarizes representative codes and highlights their treatment of defects, transport capabilities, and applicability to 2D materials.

\begin{table*}[t]
\centering
\caption{Representative computational codes for electron-defect and impurity-scattering calculations. The approaches are broadly classified into explicit first-principles treatments, in which the defect perturbation is obtained from defective and pristine DFT supercells, and model-based approaches, in which impurity scattering is described using analytical or screened interaction potentials combined with first-principles electronic structures.}
\label{tab:ed_codes}
\renewcommand{\arraystretch}{1.15}
\small
\begin{tabularx}{\textwidth}{
    >{\raggedright\arraybackslash}p{1.8cm}
    >{\raggedright\arraybackslash}p{3cm}
    >{\raggedright\arraybackslash}p{3cm}
    >{\centering\arraybackslash}p{0.95cm}
    >{\raggedright\arraybackslash}p{2.8cm}
    >{\raggedright\arraybackslash}X}
\toprule
\textbf{Code} &
\textbf{Defect treatment} &
\textbf{Defect supported} &
\textbf{2D} &
\textbf{DFT interface}&
\textbf{Main capabilities} \\
\midrule

Perturbo-ED \cite{zhou2021perturbo}
&
Explicit first-principles defect supercell
&
Neutral point defect and charged defect
&
Partial
&
Quantum ESPRESSO + Wannier90
&
Wannier interpolation, scattering rates, relaxation times, and carrier transport
\\
EDI \cite{xiao2024point}
&
Explicit first-principles defect supercell
&
Primarily neutral point defect
&
Yes
&
Quantum ESPRESSO + Wannier90
&
Wannier interpolation, mobility and conductivity, SOC supported
\\
EPW \cite{ponce2016epw}
&
Model-based charged-impurity scattering
&
Ionized impurities defect
&
Partial
&
Quantum ESPRESSO + Wannier90
&
Impurity and electron-phonon scattering, iterative Boltzmann transport
\\
QR$^2$-code \cite{huang2026QR2}
&
Constant approximation and first-principles
defect supercell 
&
Neutral point defect
&
Yes
&
Quantum ESPRESSO + EPW + Wannier90 + EDI
&
Wannier interpolation, first- and second-order resonance Raman spectra 
\\
AMSET \cite{ganose2021efficient}
&
First-principles-informed analytical
scattering models
&
Ionized impurities defects
&
No$^{a}$
&
Currently primarily VASP
&
Efficient mobility and thermoelectric transport calculations
\\
elphbolt \cite{protik2022elphbolt}
&
Model-based charged-impurity scattering
&
Ionized impurities defects
&
Yes$^{b}$
&
Quantum ESPRESSO + EPW + Wannier90
&
Coupled electron-phonon and impurity transport
\\

\bottomrule
\end{tabularx}

\vspace{1mm}
\begin{flushleft}
\footnotesize
$^{a}$Not specifically formulated for rigorous 2D scattering. 
$^{b}$2D transport is supported, while impurity scattering is described using an effective screened-Coulomb model.
\end{flushleft}
\end{table*}

As summarized in Table~\ref{tab:ed_codes}, the applicability of current electron-defect transport codes to 2D materials differs considerably. EDI~\cite{xiao2024point} provides one of the most direct first-principles frameworks for 2D point-defect scattering. In particular, the code explicitly accounts for the two-dimensional Brillouin zone, employs vacuum-level alignment of the pristine and defective supercell potentials, and expresses both carrier and defect concentrations in units of cm$^{-2}$. Its implementation has been demonstrated for defect-limited carrier transport in monolayer transition-metal dichalcogenides~\cite{xiao2024point}. 

Perturbo~\cite{zhou2021perturbo} provides a well-developed infrastructure for 2D electronic structures, Wannier interpolation, and carrier transport. However, some parts of its present electron-defect implementation, particularly the long-range treatment of charged defects, retain restrictions inherited from bulk formulations~\cite{lu2022first}; for example, the current screening implementation for charged defects is limited to cubic systems. It is therefore classified here as having partial support for 2D electron-defect calculations. Similarly, EPW~\cite{ponce2016epw} has sophisticated 2D capabilities for electron-phonon interactions, including dedicated treatments of long-range interactions in two dimensions, but its ionized-impurity module remains primarily formulated using bulk-like screened-Coulomb scattering and is not yet equivalent to a fully 2D defect-specific treatment~\cite{leveillee2023ab}.

QR$^2$-code~\cite{huang2026QR2} combines both the EPW and EDI packages. This code uses electron-defect and electron-phonon interactions—derived from EDI and EPW, respectively — as input for calculating resonance Raman spectra~\cite{liu2026defectRaman}; it also offers an option to compute Raman spectra using a simplified approximation, in which the electron-defect matrix element is treated as a constant.

The situation is different for AMSET~\cite{ganose2021efficient}. Although a 2D slab electronic structure can formally be supplied to the code, its developers have noted that the scattering matrix elements implemented in AMSET are formulated for three-dimensional systems and have different forms in two dimensions. Its quantitative application to genuine 2D carrier scattering is therefore not generally recommended without additional modifications. In contrast, elphbolt~\cite{protik2022elphbolt} contains an explicit 2D option and has been applied to several monolayer systems, although its charged-impurity scattering is likewise represented by an effective screened-Coulomb model. Consequently, for microscopic studies of specific point defects in 2D materials, EDI currently provides the most explicit defect-specific framework among the codes listed here, whereas the other packages are particularly useful when defect or impurity scattering must be combined with additional transport mechanisms.

\section{Impact of point defects on functional properties}

\subsection{Defect-limited carrier transport}

Having established the microscopic description of electron--defect scattering in Secs.~3 and 4, we now discuss how these interactions determine macroscopic transport properties, particularly electrical conductivity and carrier mobility. In the diffusive transport regime, 
these quantities can be evaluated within the Boltzmann transport equation (BTE) formalism \citep{lu2022first,claes2025phonon}. For a spatially homogeneous system under a weak electric field $\bm{\mathcal{E}}$, the steady-state linearized BTE for elastic electron-defect scattering in the dilute limit can be written as~\cite{lundstrom2010fundamentals}
\begin{equation}
    -e\bm{\mathcal{E}}\cdot\bm{v}_{n\bm{k}}
    \frac{\partial f^0_{n\bm{k}}}{\partial E_{n\bm{k}}}
    =
    \frac{n_{\mathrm{at}}C_d}{N_{\bm{k}'}}
    \sum_{m\bm{k}'}
    W^{\mathrm{def}}_{n\bm{k}\rightarrow m\bm{k}'}
    (\delta f_{m\bm{k}'}-\delta f_{n\bm{k}}),
    \label{eq:def_lbte}
\end{equation}
where $e$ is the elementary charge, $f^0_{n\bm{k}}$ is the equilibrium Fermi--Dirac distribution, $\delta f_{n\bm{k}}$ is the deviation from equilibrium, and $\bm{v}_{n\bm{k}}$ is the group velocity of the electronic state $|n\bm{k}\rangle$. The transition rate $W^{\mathrm{def}}_{n\bm{k}\rightarrow m\bm{k}'}$ is given by Eq.~\eqref{eq:def_general_fgr}. 

We introduce the vector mean free path $\bm{F}_{n\bm{k}}$, 
which describes the linear response of the electronic 
occupation to an applied electric field 
\citep{lu2022first}, as
\begin{equation}
    \delta f_{n\bm{k}} =
    e\bm{\mathcal{E}}\cdot\bm{F}_{n\bm{k}}
    \frac{\partial f^0_{n\bm{k}}}{\partial E_{n\bm{k}}}.
    \label{eq:def_occupation_change}
\end{equation}
By substituting Eq.~\eqref{eq:def_occupation_change} 
into Eq.~\eqref{eq:def_lbte}, we obtain
\begin{equation}
    \bm{v}_{n\bm{k}} =
    \frac{n_{\mathrm{at}}C_d}{N_{\bm{k}'}}
    \sum_{m\bm{k}'}
    W^{\mathrm{def}}_{n\bm{k}\rightarrow m\bm{k}'}
    \left(
    \bm{F}_{n\bm{k}}-\bm{F}_{m\bm{k}'}
    \right).
    \label{eq:def_bte_mfp}
\end{equation}
Several approaches can be used to solve  Eq.~\eqref{eq:def_bte_mfp}, including the self-energy relaxation-time approximation (SERTA) and the iterative solution of the Boltzmann transport equation (IBTE), which are implemented in Perturbo~\cite{zhou2021perturbo}, EDI~\cite{xiao2024point}, or EPW~\cite{ponce2016epw}.

Once $\bm{F}_{n\bm{k}}$ is obtained, the transport 
distribution function (TDF) for a 2D material can be 
constructed as \citep{lu2022first}
\begin{equation}
    \Sigma_{\alpha\beta}(E) =
    \frac{g_s}{N_{\bm{k}}A_{\mathrm{uc}}}
    \sum_{n\bm{k}}
    v_{n\bm{k}}^{\alpha}
    F_{n\bm{k}}^{\beta}
    \delta(E-E_{n\bm{k}}),
    \label{eq:def_tdf}
\end{equation}
where $g_s$ is the spin degeneracy, $N_{\bm{k}}$ is the 
number of $\bm{k}$ points in the Brillouin zone, and 
$A_{\mathrm{uc}}$ is the area of the primitive unit cell. 
The indices $\alpha$ and $\beta$ denote the in-plane 
Cartesian directions. For elastic scattering, the 
electrical and thermoelectric transport coefficients 
can be evaluated from the transport moments 
\citep{ricci2017ab},
\begin{equation}
    \mathcal{L}_{\alpha\beta}^{(j)} =
    \int \Sigma_{\alpha\beta}(E)
    (E-E_F)^j
    \left(-\frac{\partial f^0}{\partial E}\right)
    \mathrm{d}E,
    \label{eq:def_transport_moments}
\end{equation}
where $E_F$ is the chemical potential at temperature $T$ 
and $j$ denotes the order of the transport moment. 
The electrical conductivity tensor is given by~\cite{hung2021origin}
\begin{equation}
    \bm{\sigma} = e^2\bm{\mathcal{L}}^{(0)},
    \label{eq:def_conductivity}
\end{equation}
and the corresponding carrier mobility tensor is
\begin{equation}
    \bm{\mu} =
    \frac{\bm{\sigma}}{en_c^{\mathrm{2D}}},
    \label{eq:def_mobility}
\end{equation}
where $n_c^{\mathrm{2D}}$ is the areal concentration 
of the carrier type under consideration, with 
$\bm{\sigma}$ evaluated for the corresponding bands. 


Within the SERTA, Eqs.~\eqref{eq:def_tdf}--\eqref{eq:def_mobility}
give the defect-limited mobility as
\begin{equation}
    \mu_{\alpha\alpha}^{\mathrm{e-d}}
    =
    \frac{eg_s}{n_c^{\mathrm{2D}}N_{\bm{k}}A_{\mathrm{uc}}}
    \sum_{n\bm{k}}
    \frac{(v_{n\bm{k}}^\alpha)^2}
    {\Gamma_{n\bm{k}}^{\mathrm{e-d}}}
    \left(
    -\frac{\partial f^0_{n\bm{k}}}
    {\partial E_{n\bm{k}}}
    \right),
    \label{eq:def_mobility_microscopic}
\end{equation}
where $\Gamma_{n\bm{k}}^{\mathrm{e-d}} =1/\tau_{n\bm{k}}^{\mathrm{e-d}}$ is the scattering rate given by Eq.~\eqref{eq:40}. This expression directly connects the e-d matrix elements with the macroscopic carrier mobility. For a given electronic structure and defect concentration, weaker e-d coupling leads to lower scattering rates and consequently higher carrier mobility. Furthermore, in the dilute-defect limit, the defect-limited mobility is inversely proportional to the defect concentration, $\mu^{\mathrm{e-d}}\propto C_d^{-1}$.

Assuming independent scattering mechanisms, the combined effects of electron-phonon (e-ph) and e-d scattering on carrier mobility can be evaluated in two ways. The first is to combine the individual mobilities using Matthiessen's rule,
\begin{equation}
    \frac{1}{\mu_{\alpha\alpha}^{\mathrm{tot}}}
    \approx
    \frac{1}{\mu_{\alpha\alpha}^{\mathrm{e-ph}}}
    +
    \frac{1}{\mu_{\alpha\alpha}^{\mathrm{e-d}}},
    \label{eq:def_matthiessen_mobility}
\end{equation}
while the second is to incorporate both scattering mechanisms into the same BTE. Following Lu et al. \citep{lu2022first}, Eq.~\eqref{eq:def_bte_mfp} can be extended to include e-ph scattering as
\begin{equation}
    \begin{aligned}
        \bm{F}_{n\bm{k}}
        \left(
        \Gamma_{n\bm{k}}^{\mathrm{e-ph}}
        +
        \Gamma_{n\bm{k}}^{\mathrm{e-d}}
        \right)
        &=
        \bm{v}_{n\bm{k}}
        \\
        &+
        \frac{1}{N_{\bm{q}}}
        \sum_{m\nu\bm{q}}
        W_{n\bm{k}\rightarrow m,\bm{k}+\bm{q}}^{\nu,\mathrm{e-ph}}
        \bm{F}_{m,\bm{k}+\bm{q}}
        \\
        &+
        \frac{n_{\mathrm{at}}C_d}{N_{\bm{k}'}}
        \sum_{m\bm{k}'}
        W^{\mathrm{def}}_{n\bm{k}\rightarrow m\bm{k}'}
        \bm{F}_{m\bm{k}'}.
    \end{aligned}
    \label{eq:def_combined_collision}
\end{equation}
Here, $\Gamma_{n\bm{k}}^{\mathrm{e-ph}}$ and $\Gamma_{n\bm{k}}^{\mathrm{e-d}}$ are the total e-ph and e-d scattering rates for the electronic state $|n\bm{k}\rangle$, respectively, given by
\begin{equation}
    \Gamma_{n\bm{k}}^{\mathrm{e-ph}}
    =
    \frac{1}{N_{\bm{q}}}
    \sum_{m\nu\bm{q}}
    W_{n\bm{k}\rightarrow m,\bm{k}+\bm{q}}^{\nu,\mathrm{e-ph}},
    \label{eq:def_eph_rate}
\end{equation}
and
\begin{equation}
    \Gamma_{n\bm{k}}^{\mathrm{e-d}}
    =
    \frac{n_{\mathrm{at}}C_d}{N_{\bm{k}'}}
    \sum_{m\bm{k}'}
    W^{\mathrm{def}}_{n\bm{k}\rightarrow m\bm{k}'}.
    \label{eq:def_ed_rate}
\end{equation}
The quantity $W_{n\bm{k}\rightarrow m,\bm{k}+\bm{q}}^{\nu,\mathrm{e-ph}}$ denotes the linearized electron--phonon scattering kernel associated with the emission or absorption of a phonon with branch index $\nu$ and wavevector $\bm{q}$, including the equilibrium occupation factors \citep{zhou2021perturbo}. Once $\bm{F}_{n\bm{k}}$ is obtained by solving Eq.~\eqref{eq:def_combined_collision}, the total carrier mobility can be evaluated using Eqs.~\eqref{eq:def_tdf}--\eqref{eq:def_mobility}.

The constant relaxation-time approximation (CRTA) can be justified in the limiting case of an isotropic 2D parabolic band scattered by an unscreened short-range defect potential, as demonstrated in Eq.~\eqref{eq:def2d6}. In realistic materials, however, the momentum and band dependence of the e-d matrix elements generally requires state-dependent relaxation times. When forward scattering is significant, the momentum relaxation-time approximation (MRTA) provides a more appropriate description of momentum relaxation, whereas the iterative solution of the BTE is preferable for systems with strongly anisotropic or complex multivalley scattering.

\begin{figure*}
    \centering
    \includegraphics[width=1\linewidth]{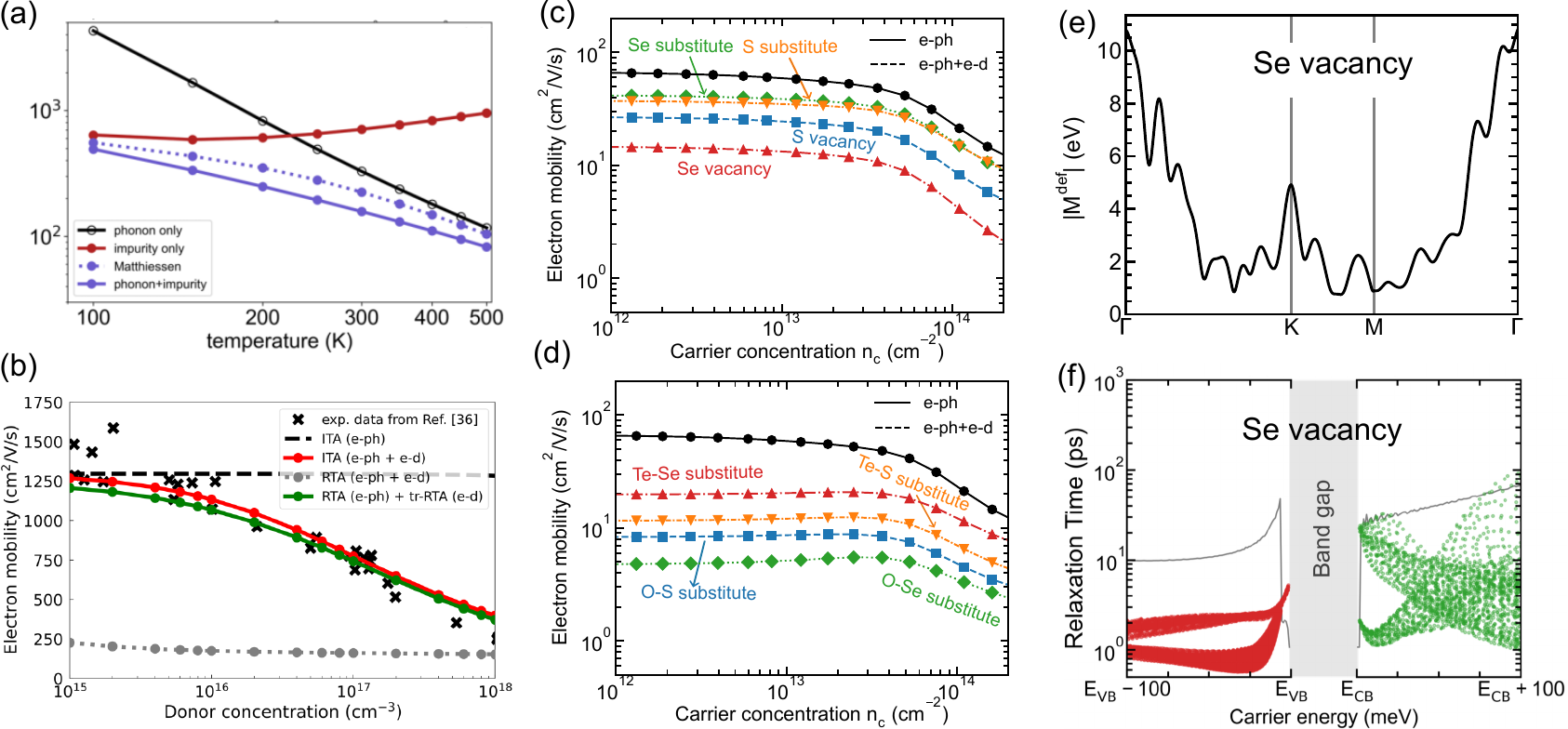}
    \caption{Impact of electron--defect scattering on carrier mobility.
    (a) Comparison of Matthiessen's rule with the combined treatment 
    of electron--phonon and ionized-impurity scattering in GaP.
    (b) Electron mobility in P-doped Si calculated using different 
    approximations to the BTE.
    (c, d) Combined phonon- and defect-limited electron mobilities 
    in Janus MoSSe as functions of carrier concentration for native 
    defects and extrinsic substitutions, respectively.
    (e) Electron-defect matrix elements magnitude $|\mathcal{M}^{\text{def}}|$ along the symmetry path for Se vacancy in Janus MoSSe. 
    (f) Semi-logarithmic plot of the energy-dependent relaxation time $\tau^\text{e-d}(E)$ in MoSSe for a Se-vacancy defect, showing holes (red dots; referenced to maximum valence band, $E_{\mathrm{VB}}$) and electrons (green dots; referenced to minimum conduction band, $E_{\mathrm{CB}}$). Thin gray curves represent the electronic DOS, scaled to arbitrary units for visual comparison with $\tau^\text{e-d}(E)$.
    (a) Reproduced from Ref.~\cite{leveillee2023ab} with permission.
    Copyright 2023 American Physical Society.
    (b) Reproduced from Ref.~\cite{lu2022first} with permission.
    Copyright 2022 American Physical Society.
    (c-f) Reproduced from Ref.~\cite{bao2026point}.
    Copyright 2026 American Physical Society.}
    \label{fig:defect_mobility}
\end{figure*}

The importance of including e-d scattering in carrier mobility calculations has been demonstrated in several bulk semiconductors. As shown in Fig.~\ref{fig:defect_mobility}(a), Leveillee et al. \citep{leveillee2023ab} found that neglecting ionized-impurity scattering substantially overestimates the electron mobility of GaP, particularly at low temperatures. Even when both scattering mechanisms are considered, Matthiessen's rule tends to overestimate the mobility obtained by incorporating both mechanisms into the same BTE. This discrepancy is particularly pronounced when electron--phonon and electron--impurity scattering contribute comparably to the reduction in mobility.

The choice of approximation for solving the BTE also affects the accuracy of the calculated mobility. Lu et al. \citep{lu2022first} investigated carrier transport in phosphorus- and boron-doped Si by incorporating both e-ph and e-d scattering into the BTE. As shown in Fig.~\ref{fig:defect_mobility}(b), the phonon-limited mobility substantially overestimates the experimental electron mobility at high donor concentrations, whereas the RTA including both scattering mechanisms significantly underestimates it. Accounting for momentum relaxation in electron-defect scattering (MRTA) brings the calculated mobility into close agreement with both the iterative solution of the BTE and experimental data.

In 2D materials, the microscopic nature of individual defects and their coupling to electronic states are important factors determining carrier mobility. Xiao et al. \citep{xiao2024point} systematically investigated the carrier mobility limited by chalcogen vacancies ($V_{\mathrm{X}}$), oxygen substitutions ($\mathrm{O}_{\mathrm{X}}$), and metal antisites ($\mathrm{M}_{\mathrm{X}}$) in monolayer MX$_2$ (M = Mo, W; X = S, Se). The electron-defect matrix elements are evaluated from first-principles defect perturbation potentials, and the corresponding mobilities are obtained within the MRTA. At a defect concentration of $10^{12}$~cm$^{-2}$, the defect-limited mobilities range from 5.8~cm$^2$V$^{-1}$s$^{-1}$ for electrons scattered by metal antisites in MoS$_2$ to 858.2~cm$^2$V$^{-1}$s$^{-1}$ for electrons scattered by oxygen substitutions in WS$_2$.

Interestingly, for the same chalcogen element, WX$_2$ consistently exhibits higher defect-limited mobilities than MoX$_2$, irrespective of the defect and carrier types considered. By separating the mobility into a Drude effective mass and a Drude relaxation time, the authors found that the longer relaxation times of WX$_2$ account for most of this difference. 
Further analysis revealed that weaker electron-defect coupling, rather than differences in the available scattering phase space, is the dominant factor responsible for the higher defect-limited mobilities of WX$_2$. These results demonstrate that the electronic band structure alone is insufficient to assess the defect-limited carrier mobility of a material.

The dependence of carrier mobility on defect species becomes particularly important in Janus TMDCs, where the two chalcogen sublayers are chemically inequivalent. Bao et al. \citep{bao2026point} investigated the combined effects of el-ph and e-d scattering in monolayer Janus MoSSe. As shown in Fig.~\ref{fig:defect_mobility}(c), at the same defect concentration of $C_d=10^{-3}$, chalcogen vacancies lead to a substantially larger reduction in electron mobility than substitutions between the native chalcogen species. Moreover, the relative suppression of mobility remains similar over a broad range of carrier concentrations.

Interestingly, although oxygen substitution improves defect-limited mobility relative to chalcogen vacancies in conventional TMDCs \citep{xiao2024point}, it leads to stronger mobility degradation in Janus MoSSe, as shown in Fig.~\ref{fig:defect_mobility}(d). This contrast highlights the dependence of electron-defect coupling on both the chemical identity of the defect and its local atomic environment. 

Figures~\ref{fig:defect_mobility}(e) and (f) show the magnitude of the electron-defect matrix element $|\mathcal{M}^{\text{def}}|$ and the energy-dependent relaxation time $\tau^{\text{e-d}}(E)=1/\Gamma^{\mathrm{e-d}}(E)$ in the Janus MoSSe monolayer with an Se-vacancy defect. It indicates that $\tau^{\text{e-d}}$ is shorter in regions with a large density of states (DOS) and longer in low-DOS regions. Consequently, the band-structure-projected $\tau^{\text{e-d}}$ exhibits pronounced minima at energies corresponding to DOS peaks.

\subsection{Quantifying defect tolerance in carrier transport}

Although defect-limited mobility provides a measure of the strength of electron-defect scattering, a more practical question for material synthesis and device fabrication is what defect concentration can be tolerated before the carrier mobility is substantially degraded.

In metals and semiconductors exhibiting metallic transport behavior, the residual resistivity ratio (RRR), defined as
\begin{equation}
    \mathrm{RRR}
    =
    \frac{\rho(T_{\mathrm{high}})}
    {\rho(T_{\mathrm{low}})},
    \label{eq:def_rrr}
\end{equation}
is widely used as an indicator of sample purity and crystalline quality \citep{fellmuth2021residual, PhysRevB.97.241203}. Here, $T_{\mathrm{low}}$ is sufficiently low that phonon scattering is strongly suppressed, and the resistivity is predominantly determined by static disorder. However, the RRR characterizes the overall residual scattering in a sample and does not directly identify the tolerable concentration of individual defect species at the operating temperature of a device.

For 2D semiconductors, Ma and Jena \citep{PhysRevX.4.011043} evaluated the critical charged-impurity density in monolayer MoS$_2$ using a screened Coulomb potential that accounts for dielectric mismatch and free-carrier screening. The impurity-limited mobility is calculated within the MRTA, while the phonon-limited mobility includes scattering by both intrinsic phonons and surface optical phonons of the surrounding dielectrics. More recently, Xiao et al. \citep{xiao2024point} extended the critical-concentration criterion to chemically distinct point defects using first-principles e-d scattering calculations. The critical defect concentration, denoted here as $C_{\mathrm{cross}}$, is determined by the condition
\begin{equation}
    \mu^{\mathrm{e-d}}(C_{\mathrm{cross}})
    =
    \mu^{\mathrm{e-ph}}.
    \label{eq:def_cross_concentration}
\end{equation}
Below this concentration, phonons provide the stronger limitation to carrier transport, whereas above it, defect scattering becomes more important.

\begin{figure}
    \centering
    \includegraphics[width=0.9\linewidth]{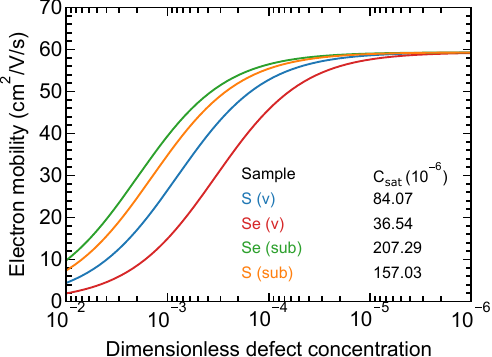}
    \caption{Defect tolerance of electron mobility in monolayer TMDCs.
    (a) Critical areal concentrations of chalcogen vacancies 
    ($V_{\mathrm{X}}$) at which the defect-limited mobility equals 
    the phonon-limited mobility in MoS$_2$, WS$_2$, MoSe$_2$, and 
    WSe$_2$.
    (b) Combined phonon- and defect-limited electron mobilities as 
    functions of defect concentration for native chalcogen-site 
    defects in monolayer Janus MoSSe at 300~K and a carrier 
    concentration of $1.2\times10^{13}$~cm$^{-2}$, together with 
    the corresponding saturation defect concentrations.
    (a) Reproduced from Ref.~\cite{xiao2024point} with permission.
    Copyright 2024 American Chemical Society.
    (b) Reproduced from Ref.~\cite{bao2026point}.
    Copyright 2026 American Physical Society.}
    \label{fig:defect-concentration}
\end{figure}

Xiao et al. \citep{xiao2024point} reported that the critical areal concentrations of chalcogen vacancies for electron transport are $2.1\times10^{11}$, $6.3\times10^{11}$, $1.6\times10^{11}$, and $7.4\times10^{11}$~cm$^{-2}$ for MoS$_2$, WS$_2$, MoSe$_2$, and WSe$_2$, respectively. For the same chalcogen element, WX$_2$ can tolerate a higher vacancy concentration than MoX$_2$ before vacancy scattering becomes the dominant mobility-limiting mechanism. Considering all the defect species and carrier types investigated by Xiao et al. \citep{xiao2024point}, the critical areal concentrations range from $4.2\times10^{10}$ to $4.5\times10^{12}$~cm$^{-2}$, demonstrating that the critical defect concentration depends strongly on both the host material and the defect species.

However, within Matthiessen's rule, the total mobility at this crossover concentration is only one-half of the phonon-limited mobility. Therefore, a more stringent criterion is required when the objective is to preserve a specified fraction of the intrinsic mobility.

More recently, Bao et al. \citep{bao2026point} introduced the saturation defect concentration, $C_{\mathrm{sat}}$, defined as the highest defect concentration for which the total mobility remains at least 90\% of its phonon-limited value. Here, a mobility-based defect-tolerance criterion can be defined as
\begin{equation}
    \mu^{\mathrm{tot}}(C_\eta)
    =
    \eta\mu^{\mathrm{e-ph}},
    \qquad 0<\eta<1,
    \label{eq:def_tolerance1}
\end{equation}
where $\eta$ represents the fraction of the phonon-limited mobility to be retained.
On the other hand, in the dilute-defect limit, the defect-limited mobility scales as
\begin{equation}
    \mu^{\mathrm{e-d}}(C_d)
    =
    \frac{C_{\mathrm{ref}}}{C_d}
    \mu^{\mathrm{e-d}}(C_{\mathrm{ref}}),
\label{eq:def_tolerance2}
\end{equation}
where $C_{\mathrm{ref}}$ is an arbitrary reference defect concentration within the dilute-defect regime. By substituting Eqs.~\eqref{eq:def_tolerance1} and~\eqref{eq:def_tolerance2} into Eq.~\eqref{eq:def_matthiessen_mobility} yields
\begin{equation}
    C_\eta
    =
    C_{\mathrm{ref}}
    \frac{\mu^{\mathrm{e-d}}(C_{\mathrm{ref}})}
    {\mu^{\mathrm{e-ph}}}
    \left(\frac{1-\eta}{\eta}\right).
    \label{eq:def_tolerance_concentration}
\end{equation}
Consequently, the saturation defect concentration can be defined as follows \cite{bao2026point}:
\begin{equation}
    C_{\mathrm{sat}}
    =
    \frac{C_{\mathrm{ref}}}{9}
    \frac{\mu^{\mathrm{e-d}}(C_{\mathrm{ref}})}
    {\mu^{\mathrm{e-ph}}}.
    \label{eq:def_csat}
\end{equation}
This gives $C_{\mathrm{sat}}=C_{\mathrm{cross}}/9$ within the dilute-defect limit, assuming independent scattering mechanisms and the validity of Matthiessen's rule.


In Fig.~\ref{fig:defect-concentration}, we show how electron mobility evolves with the concentration of different defect species in monolayer Janus MoSSe.
As the defect concentration decreases from $10^{-2}$ to $10^{-6}$, the electron mobility gradually approaches its phonon-limited value of approximately 59.34~cm$^2$V$^{-1}$s$^{-1}$ at 300~K. However, the defect concentration required to retain 90\% of this mobility differs considerably among the defect species. As shown in Fig.~\ref{fig:defect-concentration}, the saturation defect concentrations of S and Se vacancies are 84.07 and 36.54~ppm, respectively, whereas those of Se substitutions at S sites and S substitutions at Se sites are 207.29 and 157.03~ppm, respectively.
Among these four defect species, Se vacancies impose the most stringent constraint on defect concentration, while Se substitutions at S sites can be tolerated at concentrations nearly six times higher. For extrinsic substitutions, even lower saturation defect concentrations were obtained, reaching 8.31~ppm for oxygen substitutions at Se sites \cite{bao2026point}.

These results demonstrate that defect tolerance depends on both the host material and the chemical identity and atomic site of the defect. Consequently, experimental efforts to improve carrier mobility should focus on defect species with low saturation defect concentrations.

\subsection{Enhanced carrier mobility by electron cloaking}
Electron-defect interactions are generally considered detrimental to carrier transport because defects and ionized dopants perturb the electronic potential and scatter propagating carriers, as shown in Sec. 5.1. This effect becomes particularly important in heavily doped semiconductors, where a high dopant concentration is required to achieve a large carrier density, but the associated charged defects can strongly reduce the carrier mobility. Conventionally, the mobility degradation caused by ionized dopants has been attributed primarily to their long-range Coulomb fields. However, first-principles calculations have shown that the defect potential can deviate substantially from a simple Coulomb potential close to the defect, where the local chemical environment, atomic relaxation, and electronic structure become important~\cite{liao2012cloaking,liao2013isotropic,zhou2022mobility}. Importantly, these short-range contributions do not simply provide an additional independent scattering channel. Instead, they enter coherently into the electron-defect coupling matrix element together with the long-range Coulomb contribution. This provides the physical basis for \textit{electron cloaking}, in which the short-range and long-range contributions partially cancel, strongly suppressing electron-defect scattering.

\begin{figure*}[t]
  \centering
  \includegraphics[width=0.95\linewidth]{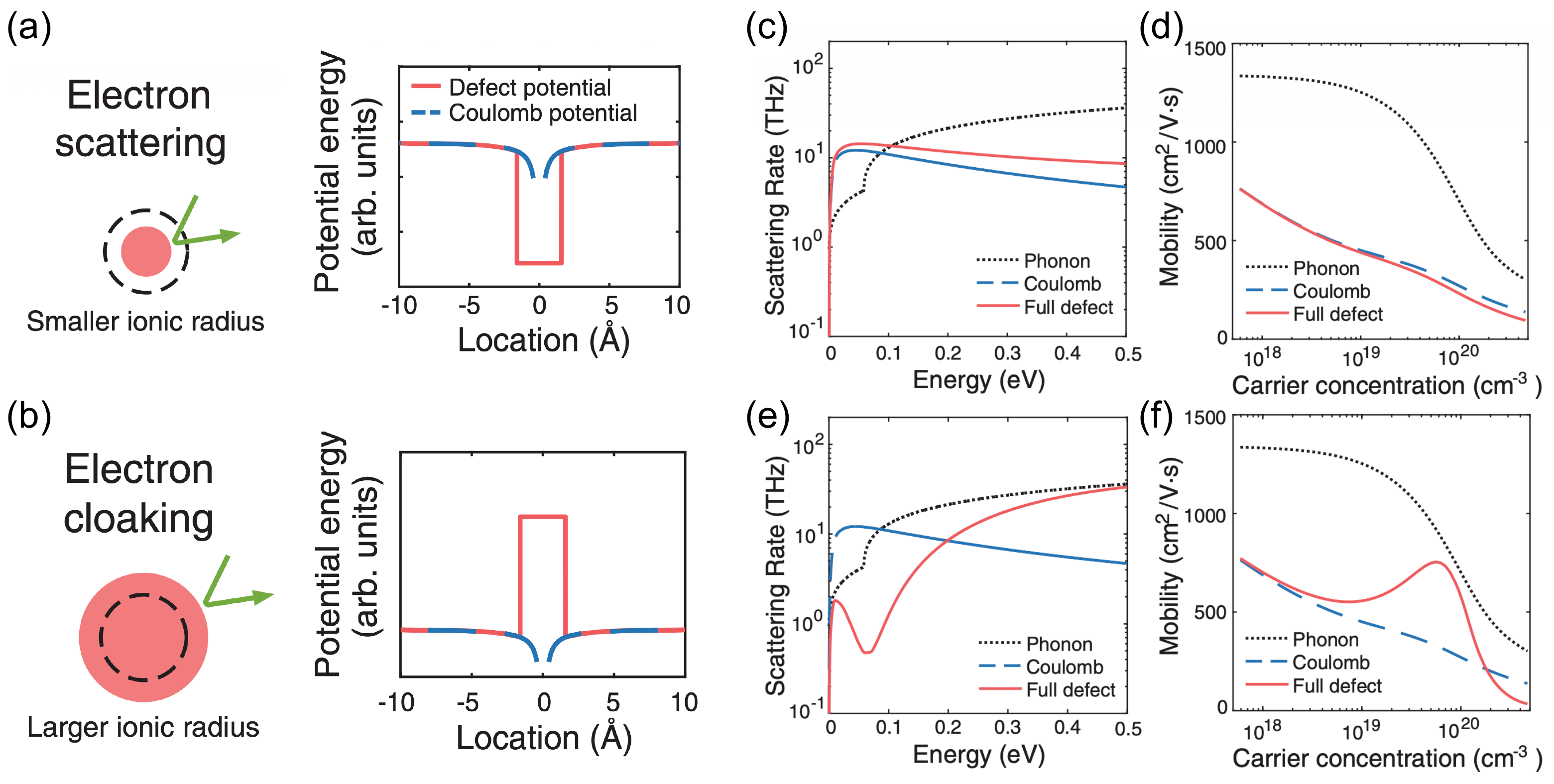}
  \caption{\label{fig:cloaking} Defect potentials corresponding to two different scenarios in an
  $n$-type semiconductor: (a) electron scattering due to a smaller ionic radius than the host atom and (b) electron cloaking due to a larger ionic radius than the host atom. (c, e) Electron scattering rates due to electron–phonon interactions, compared with those considering electron-defect scattering for the electron-scattering and electron-cloaking cases, respectively. The calculation assumes a parabolic band to illustrate the general impact of defect scattering. (d, f) Electron mobility is plotted as a function of carrier concentration for the electron-scattering and electron-cloaking cases, respectively. In Fig. (d), the mobility at high carrier concentrations is traditionally limited by Coulomb scattering. However, electron cloaking can break this limit, leading to high mobility limited only by the intrinsic electron–phonon interactions, as shown in Fig. (f). Reproduced from Ref.~\cite{zhou2022mobility} with permission. CC BY 4.0.}
\end{figure*}

For a charged-defect, the defect potential $\Delta V_{\text{def}}$ in Eq.~\eqref{eq:2} is defined by the term of the short-range $\Delta V_{\text{sr}}$ and long-range $\Delta V_{\text{lr}}$ potentials as:
\begin{equation}
\Delta \widetilde{V}_{\text{def}}(\bm q) = \Delta \widetilde{V}_{\text{sr}}(\bm q) + \Delta \widetilde{V}_{\text{lr}}(\bm q),
\label{eq:5.2.1}
\end{equation}
where the short-range part can be calculated within the supercell based on the Fourier transform, as shown in Eq.~\eqref{eq:33}, i.e., 
\begin{equation}
\Delta \widetilde{V}_{\text{sr}}(\bm q)=\frac{1}{\Omega_0}\int_{\Omega_{\mathrm{sup}}}
\Delta V_{\text{sr}}(\bm{r})
e^{-i\bm{q}\cdot\bm{r}}\mathrm{d}\bm r,
\label{eq:5.2.2}
\end{equation}
while the long-range part can be obtained by integrating
analytically to infinity, yielding~\cite{zhou2022mobility}
\begin{equation}
\begin{split}
\Delta \widetilde{V}_{\text{lr}}(\bm q)&=\frac{1}{\Omega_0}\int_{\mathbb{R}^3}
\Delta V_{\text{lr}}(\bm{r})
e^{-i\bm{q}\cdot\bm{r}}\mathrm{d}\bm r\\
&=-\frac{Ze^2}{\Omega_0\varepsilon\varepsilon_0}\frac{1}{|\bm q|^2+(1/L_D)^2},
\end{split}
\label{eq:5.2.3}
\end{equation}
in which we have assumed the Coulomb potential energy is given by $V_{\text{lr}}(\bm{r})=-\frac{Ze^2}{4\pi\varepsilon\varepsilon_0}\frac{\exp{(-r/L_D)}}{r}$, where the factor $\exp{(-r/L_D)}$ considers the carrier screening at high carrier concentrations, with the Debye screening length $L_D$ given by
\begin{equation}
 L_D=\left[\frac{e^2}{\varepsilon\varepsilon_0}\int\left(-\frac{\partial f^0}{\partial E}\mathcal{D}(E)\mathrm{d} E\right)\right]^{-1/2}.
\label{eq:5.67}
\end{equation}

By substituting Eq.~\eqref{eq:5.2.2} into Eq.~\eqref{eq:2}, the e-d matrix element $\mathcal{M}_{mn}^{\mathrm{def}}(\bm{k}',\bm{k})$ can be separated into the long-range $\mathcal{M}^{\mathrm{lr}}$ and short-range $\mathcal{M}^{\mathrm{sr}}$ contributions as follows:
\begin{equation}
\begin{split}
\mathcal{M}_{mn}^{\mathrm{def}}
(\bm{k}',\bm{k})
&=
\left\langle
m\bm{k}'
\left|
\Delta V_{\mathrm{lr}}
+
\Delta V_{\mathrm{sr}}
\right|
n\bm{k}
\right\rangle
\ \\
&=\left\langle
m\bm{k}'
\left|
\Delta V_{\mathrm{lr}}\right|
n\bm{k}
\right\rangle+\left\langle
m\bm{k}'
\left|
\Delta V_{\mathrm{sr}}
\right|
n\bm{k}
\right\rangle
\ \\
&=
\mathcal{M}_{mn}^{\mathrm{lr}}
(\bm{k}',\bm{k})
+
\mathcal{M}_{mn}^{\mathrm{sr}}
(\bm{k}',\bm{k}).
\end{split}
\label{eq:5.68}
\end{equation}
When $\left|
\mathcal{M}^{\mathrm{lr}}
+
\mathcal{M}^{\mathrm{sr}}
\right|
\rightarrow 0$, the defect becomes nearly ``invisible'' to the propagating carrier even though the defect itself remains physically present. This is the central concept of electron cloaking.

This coherent cancellation also explains why the long-range and short-range contributions should not be treated as two independent scattering mechanisms. If separate scattering rates are calculated and subsequently added, the interference term in Eq.~\eqref{eq:5.2.1} would be lost, and both contributions would necessarily increase the total scattering rate due to Matthiessen's rule. Electron cloaking can therefore only be properly described by constructing the full defect potential and evaluating the complete e-d coupling matrix element before calculating the scattering rate~\cite{zhou2022mobility}. It is noted that compared with the e-d scattering rates of a neutral defect, the e-d scattering rates of the charged defect need to consider the momentum-relaxation rate as follows:
\begin{equation}
\begin{split}
\frac{1}{\tau_{n\bm{k}}}=
\frac{2\pi}{\hbar}\frac{n_{\text{at}}C_{\mathrm{d}}}{N_{\bm{k}'}}\sum_{m\bm{k}'}&\left(1-\frac{\bm{v}_{n\bm k}\cdot \bm{v}_{m\bm k'}}{|\bm{v}_{n\bm k}||\bm{v}_{m\bm k'}|}\right)\\
&\times|\mathcal{M}_{mn}^{\mathrm{def}}(\bm{k}',\bm{k})|^2\delta(E_{n\bm{k}}-E_{m\bm{k}'}),
\end{split}
\label{eq:5.2.4}
\end{equation}
where the factor $1-\frac{\bm{v}_{n\bm k}\cdot \bm{v}_{m\bm k'}}{|\bm{v}_{n\bm k}||\bm{v}_{m\bm k'}|}$ accounts for the efficiency of each scattering event in relaxing the carrier momentum. Forward scattering, for which the initial and final velocities are nearly parallel, contributes little to momentum relaxation, whereas large-angle and backscattering events contribute more strongly to electrical resistance.

For an $n$-type semiconductor, dopants with a smaller ionic radius than the host atom tend to introduce an additional attractive short-range potential, reinforcing the Coulomb interaction and enhancing e-d scattering [Fig.~\ref{fig:cloaking}(a)]. In contrast, dopants with a larger ionic radius tend to generate a repulsive short-range potential that counteracts the attractive Coulomb field, thereby realizing the electron-cloaking condition when the cancellation becomes sufficiently strong [Fig.~\ref{fig:cloaking}(b)]. The opposite trend applies to $p$-type materials, where the long-range Coulomb potential is repulsive and smaller-radius dopants, which generate attractive short-range potentials, are therefore more favorable. Correspondingly, the e-d scattering rate is enhanced in the scattering regime [Fig.~\ref{fig:cloaking}(c)] but strongly reduced under the cloaking condition [Fig.~\ref{fig:cloaking}(e)]. As a result, the carrier mobility decreases when defect scattering is reinforced [Fig.~\ref{fig:cloaking}(d)], whereas it can approach the intrinsic electron-phonon-limited value when electron cloaking suppresses defect scattering [Fig.~\ref{fig:cloaking}(f)].

For an $n$-type semiconductor, dopants with a smaller ionic radius than the host atom tend to introduce an additional attractive short-range potential, reinforcing the Coulomb interaction and enhancing e-d scattering (Fig.~\ref{fig:cloaking}(a)). In contrast, dopants with a larger ionic radius tend to generate a repulsive short-range potential that counteracts the attractive Coulomb field, thereby realizing the electron-cloaking condition when the cancellation becomes sufficiently strong (Fig.~\ref{fig:cloaking}(b)). The opposite trend applies to $p$-type materials, where the long-range Coulomb potential is repulsive and smaller-radius dopants, which generate attractive short-range potentials, are therefore more favorable. Correspondingly, the e-d scattering rate is enhanced in the scattering regime (Fig.~\ref{fig:cloaking}(c)) but strongly reduced under the cloaking condition (Fig.~\ref{fig:cloaking}(e)). As a result, the carrier mobility decreases when defect scattering is reinforced (Fig.~\ref{fig:cloaking}(d)), whereas it can approach the intrinsic electron-phonon-limited value when electron cloaking suppresses defect scattering (Fig.~\ref{fig:cloaking}(f)).

This theoretical prediction is also supported by transport data in heavily doped semiconductors and half-Heusler compounds. For example, in $p$-type NbFeSb, the short-range potential associated with selected dopants partially counteracts the long-range Coulomb interaction. This reduction in e-d scattering brings carrier transport closer to the intrinsic electron-phonon limit and improves the calculated electrical conductivity and thermoelectric power factor. These results show that the chemical identity of a dopant can be deliberately chosen to engineer the electron-defect interaction rather than merely control the carrier concentration~\cite{zhou2022mobility}.

\begin{figure}[t]
  \centering
  \includegraphics[width=1.0\linewidth]{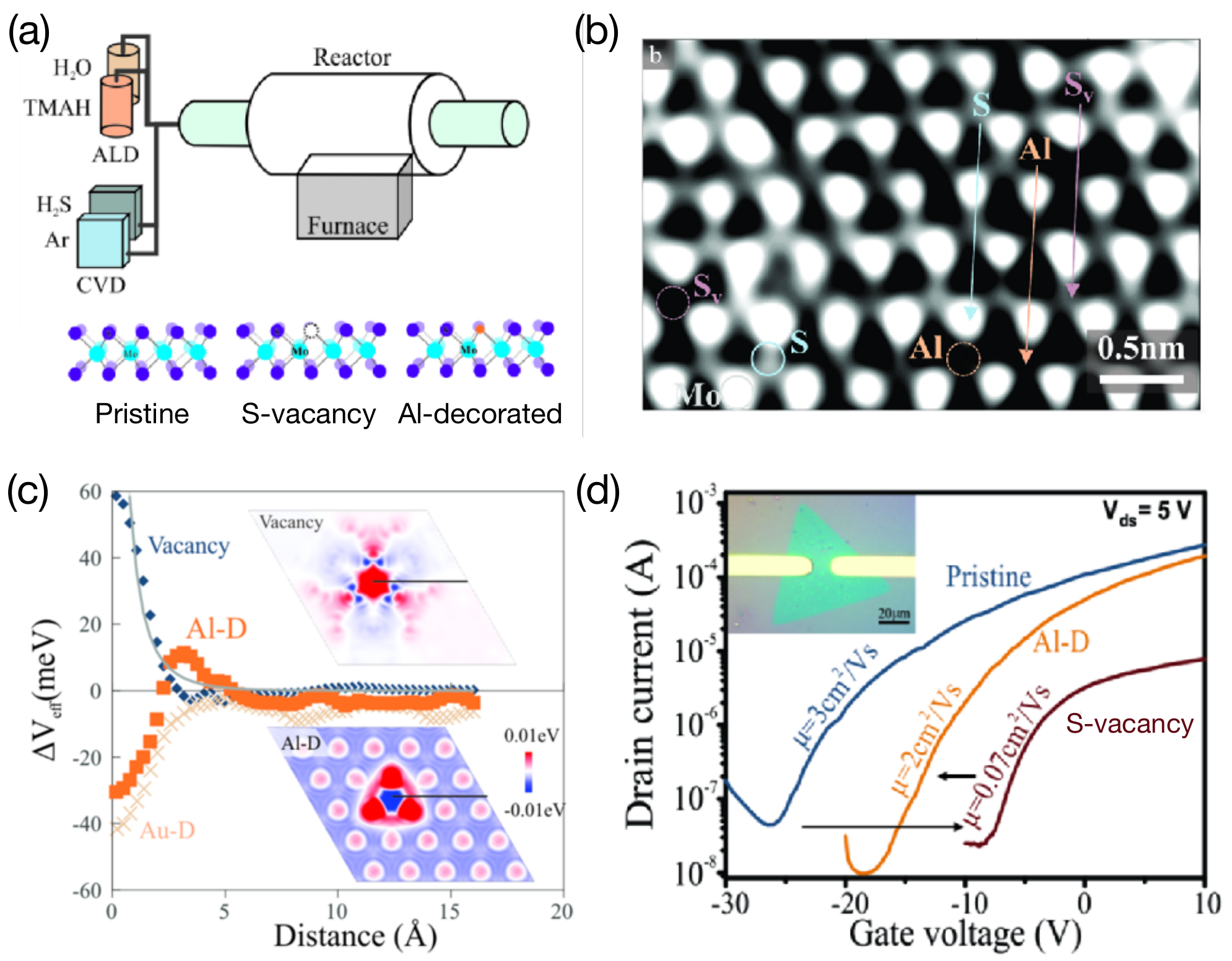}
  \caption{\label{fig:cloaking2} (a) Experimental setup for synthesizing the pristine, S-vacancy, and Al-decorated (Al-D) MoS$_2$ monolayers. (b) Annular dark-field scanning transmission electron microscopy (ADF-STEM) atomic-resolution image of Al-decorated MoS$_2$ with indication of different elements (Mo, Al, S) and S-vacancy(S$_\text{V}$). (c) Simulated effective potential from DFT calculations for S-vacancy, Au-decorated, and Al-decorated MoS$_2$, with insets showing the spatial distribution of the effective potential in the Mo planes for S-vacancy-defected and Al-decorated cases. (d) Transconductance curves for field-effect transistors (FETs) for pristine, S-vacancy, and Al-decorated (Al-D) MoS$_2$ monolayers, with the inset showing a representative optical micrograph of an MoS$_2$ FET. Reproduced from Ref.~\cite{chen2025electron} with permission. Copyright 2025 American Chemical Society.}
\end{figure}

More recently, Chen \textit{et al.}~\cite{chen2025electron} experimentally demonstrated electron cloaking in a 2D semiconductor using sulfur-vacancy-defected monolayer MoS$_2$. Sulfur vacancies are common defects in MoS$_2$ and generate a long-range Coulomb perturbation that strongly scatters conduction electrons. Instead of eliminating these vacancies, Al atoms are selectively introduced at the vacancy sites by atomic layer deposition (Figs.~\ref{fig:cloaking2}(a) and (b)). DFT calculations showed that Al decoration substantially modifies the effective defect potential (Fig.~\ref{fig:cloaking2}(c)). Whereas the bare sulfur vacancy exhibits a relatively monotonic screened Coulomb potential, the Al-decorated vacancy produces a nonmonotonic potential due to competition between the localized Al-induced interaction and the longer-range vacancy potential. This opposing local contribution reduces the overall e-d scattering strength and realizes the cloaking condition. Carrier transport measurements in Fig.~\ref{fig:cloaking2}(d) show that the pristine MoS$2$ exhibited a field-effect mobility of approximately $\approx 3~\mathrm{cm^2 V^{-1}s^{-1}}$,
whereas increasing the sulfur-vacancy concentration through annealing reduced the mobility by approximately a factor of 40. Subsequent Al decoration restored the mobility to nearly the pristine value~\cite{chen2025electron}.

This result highlights an important distinction between electron cloaking and conventional defect passivation. In defect passivation, the objective is generally to eliminate electrically active defect states or reduce the defect concentration. Electron cloaking instead aims to engineer the defect potential so that the defect remains present but couples only weakly to propagating carriers.

\subsection{Effect of electron-defect interaction on thermoelectricity}

In thermoelectric materials, point defects are often introduced to enhance phonon scattering and suppress lattice thermal conductivity. However, these defects can also scatter charge carriers and degrade electronic transport. Evaluating the overall effect of point defects on thermoelectric performance therefore requires consideration of both electronic and phononic transport properties.

The Seebeck coefficient, which characterizes the voltage generated in response to a temperature gradient, can also be obtained from the transport moments in Eq.~\eqref{eq:def_transport_moments} as~\cite{hung2021origin}:
\begin{equation}
    \bm{S} =
    -\frac{1}{eT}
    \left(\bm{\mathcal{L}}^{(0)}\right)^{-1}
    \bm{\mathcal{L}}^{(1)}.
    \label{eq:def_seebeck}
\end{equation}
Combining the electrical conductivity in Eq.~\eqref{eq:def_conductivity} and the Seebeck coefficient in Eq.~\eqref{eq:def_seebeck}, the thermoelectric power factor (PF) along a given transport direction is obtained as
\begin{equation}
    \mathrm{PF} = \sigma S^2,
    \label{eq:def_PF}
\end{equation}
which characterizes the ability of a material to generate electrical power from a temperature gradient~\cite{hung2016quantum}.

However, efficient thermoelectric energy conversion also requires low thermal conductivity to maintain the temperature difference across the material. The overall thermoelectric performance is therefore commonly evaluated using the dimensionless figure of merit, $ZT$, defined as~\cite{goldsmid2010introduction}
\begin{equation}
    ZT = \frac{\mathrm{PF}\,T}{\kappa_e + \kappa_{ph}}
       = \frac{\sigma S^2 T}{\kappa_e + \kappa_{ph}},
    \label{eq:ZT}
\end{equation}
where $\kappa_e$ and $\kappa_{ph}$ are the electronic and phononic thermal conductivities, describing heat transport by charge carriers and phonons, respectively. A high $ZT$ thus requires a combination of a high power factor and low thermal conductivity.

\begin{figure}
    \centering
    \includegraphics[width=1\linewidth]{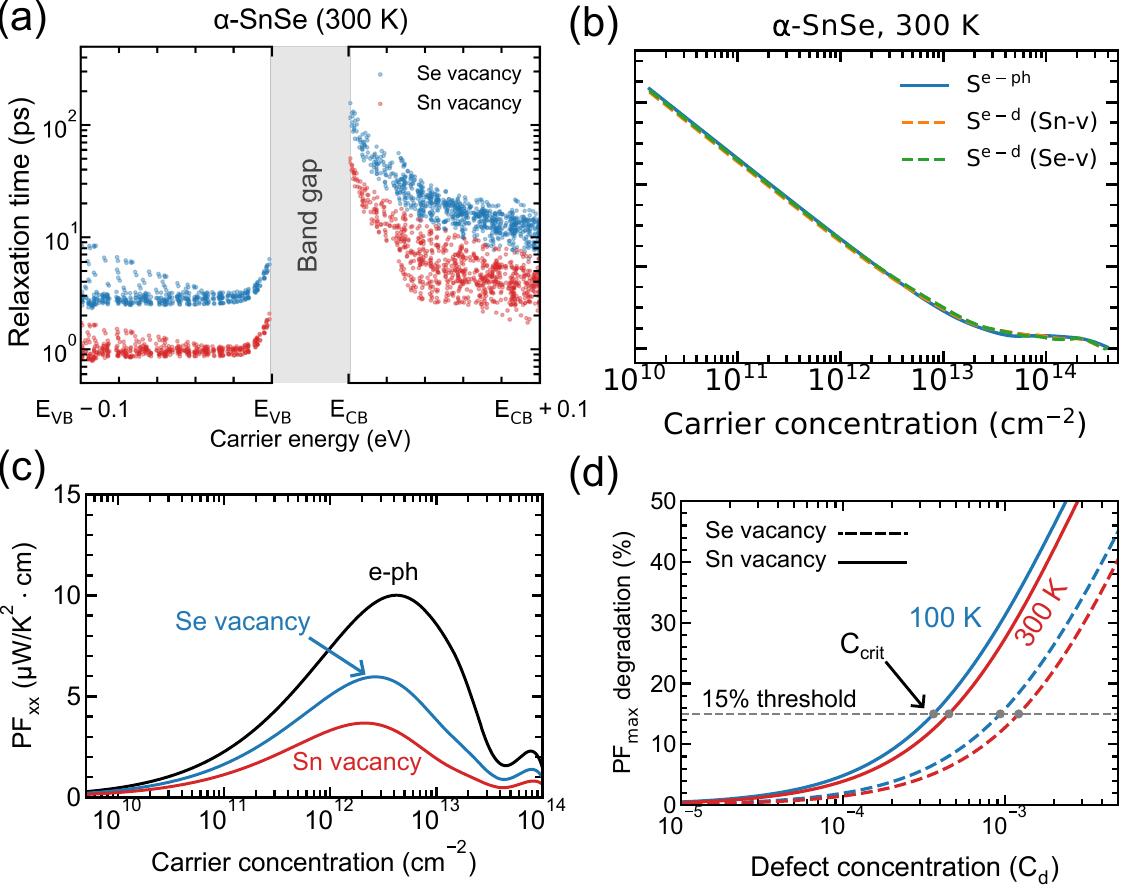}
    \caption{Effects of Sn and Se vacancies on thermoelectric transport 
    in $\alpha$-SnSe. (a) Defect-limited carrier relaxation times at 
    300~K and $C_d=10^{-6}$. (b) Seebeck coefficient and (c) power factor 
    as functions of electron concentration at 300~K, with 
    $C_d=5\times10^{-3}$ in (c). (d) Maximum power factor degradation 
    as a function of $C_d$ at 100 and 300~K, with a 15\% threshold 
    defining $C_{\mathrm{crit}}$.
    Reproduced from Ref.~\cite{bao2026anharmonic} under the 
    CC BY 4.0 license.}
    \label{fig:thermoelectric}
\end{figure}

Bao et al.~\cite{bao2026anharmonic} investigated how point-defect scattering affects the thermoelectric power factor of monolayer SnSe. The defect-specific scattering rates are evaluated from the e-d coupling matrix elements defined in Eq.~\eqref{eq:2}, following Eq.~\eqref{eq:40}. In the dilute-defect limit, these scattering rates increase linearly with the defect concentration $C_d$. For both $\alpha$- and $\beta$-SnSe, Se vacancies ($V_{\mathrm{Se}}$) produce weaker e-d scattering than Sn vacancies ($V_{\mathrm{Sn}}$), as illustrated by the longer carrier relaxation times in Fig.~\ref{fig:thermoelectric} (a). Since the electrical conductivity depends directly on the carrier relaxation times through the transport moments, stronger e-d scattering is expected to reduce the thermoelectric power factor.

Assuming independent e-ph and e-d scattering and additive resistivities, the total Seebeck coefficient was calculated using the Nordheim--Gorter rule:
\begin{equation}
    S(C_d) =
    \frac{
        \rho_{\mathrm{e-ph}}S_{\mathrm{e-ph}}
        + \rho_{\mathrm{e-d}}(C_d)S_{\mathrm{e-d}}
    }{
        \rho_{\mathrm{e-ph}}+\rho_{\mathrm{e-d}}(C_d)
    },
    \label{eq:Nordheim_Gorter}
\end{equation}
where $\rho_{\mathrm{e-ph}}$ and $\rho_{\mathrm{e-d}}(C_d)$ are the partial 
resistivities associated with e-ph and e-d scattering, respectively, and 
$S_{\mathrm{e-ph}}$ and $S_{\mathrm{e-d}}$ are their corresponding 
Seebeck coefficients.

As shown in Fig.~\ref{fig:thermoelectric}(b), the Seebeck coefficients obtained from e-ph and e-d scattering are nearly identical over the relevant carrier-concentration range. Within the dilute-defect approximation, the defect-limited relaxation times scale as $C_d^{-1}$. This common prefactor cancels in the ratio of the transport moments in Eq.~\eqref{eq:def_seebeck}, making the defect-limited Seebeck coefficient independent of $C_d$. Nevertheless, the Seebeck coefficient remains sensitive to the electronic structure and the energy dependence of the relaxation times. Given the calculated similarity between $S_{\mathrm{e-ph}}$ and $S_{\mathrm{e-d}}$, the Nordheim--Gorter rule can therefore be approximated as

\begin{equation}
    S(C_d) \approx S_{\mathrm{e-ph}}.
    \label{eq:seebeck_approx}
\end{equation}
Under this approximation, the power factor becomes
\begin{equation}
    \mathrm{PF}(C_d) \approx
    \frac{S_{\mathrm{e-ph}}^2}
    {\rho_{\mathrm{e-ph}}+\rho_{\mathrm{e-d}}(C_d)}.
    \label{eq:PF_defect}
\end{equation}
The degradation of the power factor is therefore primarily governed 
by the additional electrical resistivity induced by e-d scattering.

As shown in Fig.~\ref{fig:thermoelectric}(c), in $\alpha$-SnSe at 300~K and $C_d=5\times10^{-3}$, the maximum n-type power factor decreases from $10.014~\mu\mathrm{W}/(\mathrm{K}^{2}\,\mathrm{cm})$ in the phonon-limited case to $5.969$ and $3.681~\mu\mathrm{W}/(\mathrm{K}^{2}\,\mathrm{cm})$ for $V_{\mathrm{Se}}$ and $V_{\mathrm{Sn}}$, respectively. This difference is consistent with the defect-limited relaxation times shown in Fig.~\ref{fig:thermoelectric}(a).

To quantify defect tolerance, the authors introduced a critical defect concentration, $C_{\mathrm{crit}}$, defined as the defect concentration at which the maximum power factor decreases by 15\% relative to its phonon-limited value. As shown in Fig.~\ref{fig:thermoelectric}(d), $V_{\mathrm{Se}}$ exhibits a higher $C_{\mathrm{crit}}$ than $V_{\mathrm{Sn}}$ at both 100 and 300~K, indicating that a larger concentration of Se vacancies can be tolerated before the power factor is substantially degraded. Furthermore, stronger e-ph scattering at higher temperatures reduces the relative contribution of e-d scattering to the total resistivity, resulting in a higher $C_{\mathrm{crit}}$.

These results highlight the importance of defect-specific electron scattering in thermoelectric materials. Determining whether a particular defect improves $ZT$ requires evaluating its effects on both electronic and phononic transport. However, rather than simply limiting the concentration of defects to preserve the power factor, one may exploit electron-defect interactions to improve thermoelectric performance. As discussed in Sec.~5.2, electron cloaking provides a potential strategy for suppressing e-d scattering and maintaining high carrier mobility even at high doping concentrations~\citep{zhou2022mobility}. Furthermore, engineering the energy dependence of e-d scattering may enable selective carrier scattering to enhance the Seebeck coefficient. These strategies, combined with the design of defects that strongly scatter phonons while weakly scattering charge carriers, may help decouple electronic and phononic transport, opening opportunities for enhancing both the power factor and $ZT$ ~\citep{zhou2026material}.

\subsection{Point-defect-induced Raman spectra}

\begin{figure*}[t]
  \centering
  \includegraphics[width=0.75\linewidth]{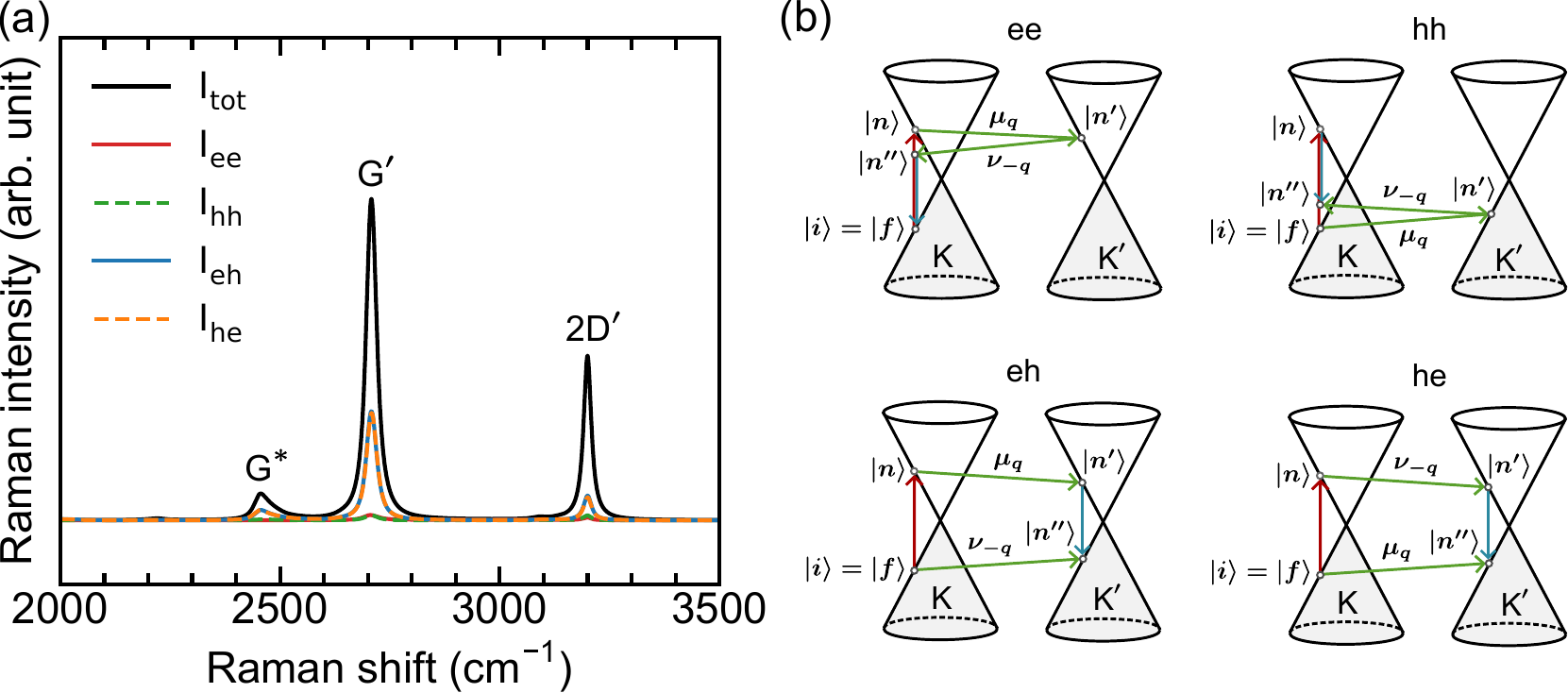}
  \caption{\label{fig:DRR1} (a) Double-resonance Raman spectra (DRR) of graphene at a laser energy of 1.96~eV, in which the three DRR peaks correspond to the G$^*$, G$^\prime$, and 2D$^\prime$ modes. (b) Schematics of the DRR scattering pathways, including $ee$, $hh$, $eh$, and $he$, with either the electron, the hole, or both involved in the carrier-phonon scattering process. Reproduced from Ref.~\citep{huang2026QR2} with permission. Copyright 2026 Elsevier.}
\end{figure*}

Raman spectroscopy is a widely used, nondestructive probe of defects in 2D materials~\cite{saito2025deep}. In graphene, the defect-activated D and D$'$ bands have been extensively studied, and their assignments and double-resonance origins are summarized in the review by Saito \textit{et al.}~\citep{saito2011raman}. The D band is associated with intervalley scattering involving a transverse-optical phonon near the K point, whereas the D$'$ band involves intravalley scattering by a longitudinal-optical phonon near the $\Gamma$ point. The double-resonance framework provides the microscopic basis for these assignments~\citep{thomsen2000double,venezuela2011}.

Figure~\ref{fig:DRR1} illustrates the double-resonance Raman (DRR) spectrum of pristine graphene and the $ee$, $hh$, $eh$, and $he$ carrier pathways. In a defect-induced DRR process, one of the phonon-scattering events is replaced by elastic scattering from a defect, which supplies the required crystal momentum. The figure therefore provides the background for the carrier pathways discussed below; it is not itself a calculated defect-induced spectrum.

Early approaches to interpreting defect-induced Raman spectra included phenomenological models such as the phonon-confinement model. Although these models can describe changes in peak positions and linewidths, they do not necessarily identify the microscopic scattering mechanisms. Tight-binding calculations based on DRR theory provided a more microscopic description of defect-activated bands in graphene~\cite{i1179,basko08}. Subsequent first-principles methods made it possible to calculate resonant Raman spectra using electronic and vibrational properties obtained from density functional theory~\cite{hung2024qeraman,saito2026theoretical}. The QR$^2$-code extended first-principles Raman calculations to defect-induced DRR spectra~\cite{huang2026QR2}. In its graphene defect tutorial, however, the elastic electron-defect matrix element was approximated by a constant and set to 1; it was not calculated for a specific defect. This approximation allows the defect-activated peak positions and scattering pathways to be illustrated, but does not provide defect-specific intensities or a quantitative comparison with the two-phonon Raman intensity~\citep{huang2026QR2}. Simplified or phenomenological descriptions of defect scattering have also been used in earlier work, although other studies explicitly calculated the defect matrix elements for selected defect models~\citep{venezuela2011}.

Recently, Liu \textit{et al.}~\citep{liu2026defectRaman} introduced
the unit-cell double-resonance Raman (Uc-DRR) and supercell single-resonance Raman (Sc-SRR) approaches, which are implemented in the updated QR$^2$-code. The Uc-DRR evaluates the electron-defect matrix element from the defect-induced perturbation potential, as discussed in Sec. 4. It is evaluated between pristine Bloch states using the perturbation potential defined in Eq.~\eqref{eq:def_deltaV}. Here, we retain only the $ee$ carrier pathway, in which both the phonon and defect scattering events act on the photoexcited electron, and omit the $ee$ superscript for brevity. We retain both possible time orderings: phonon scattering followed by defect scattering ($\mathrm{pd}$), and defect scattering followed by phonon scattering ($\mathrm{dp}$).

\begin{figure*}[t]
  \centering
  \includegraphics[width=0.85\linewidth]{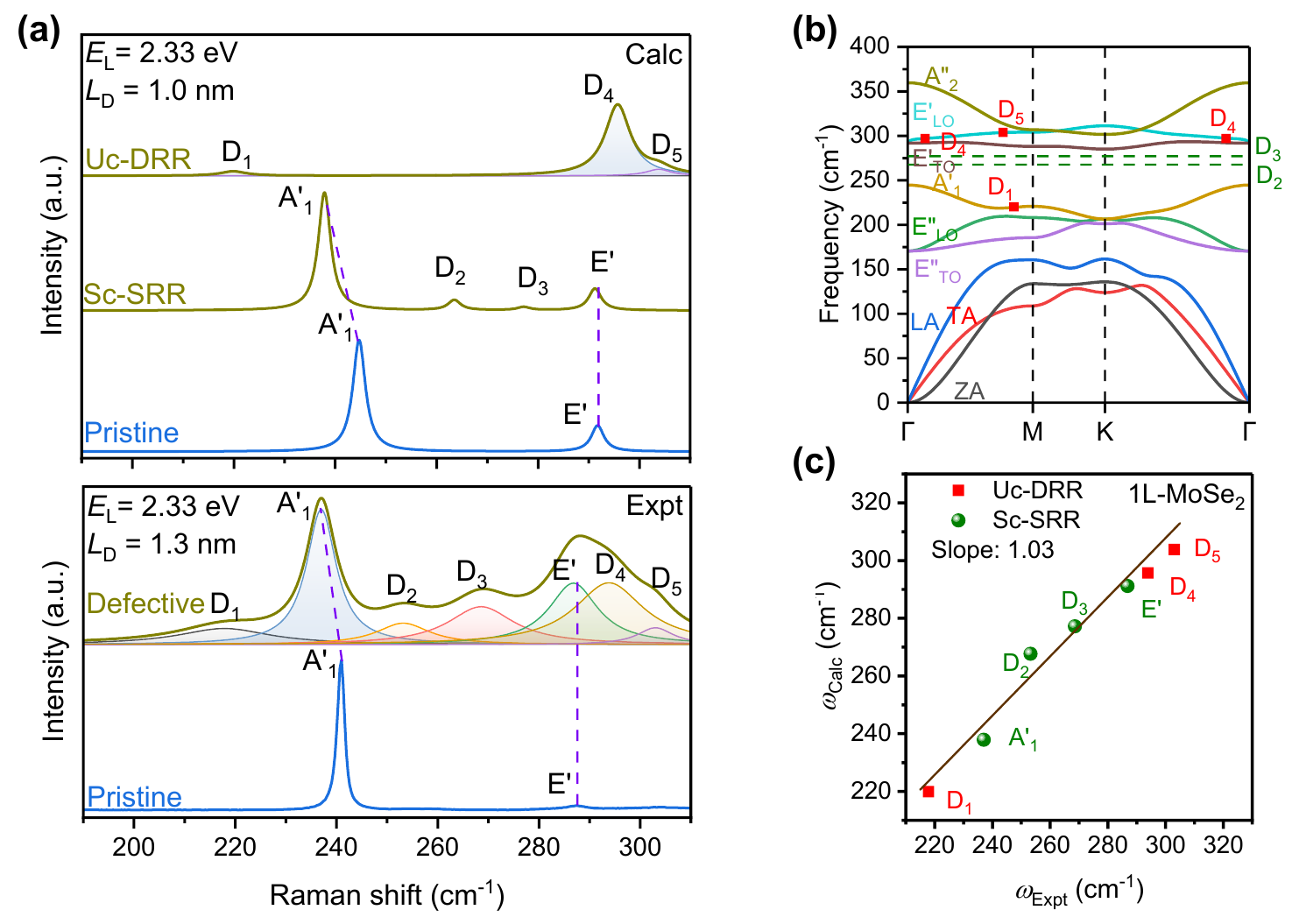}
  \caption{\label{fig:mose2}
  (a) Comparison between pristine and defective Raman spectra at $E_\mathrm{L}=\SI{2.41}{eV}$: calculation (top) and experiment~\cite{lucchese10} (bottom). 
  Both calculated and experimental defect distances are $L_\mathrm{D}=\SI{1.0}{nm}$. The calculated Raman spectra by both Uc-DRR and Sc-SRR approaches are presented in the top panel.
  (b) Peak position as a function of $E_\mathrm{L}$ for D (top) and $\rm D^\prime$ (bottom) bands. The red squares and green circles represent calculated and experimental~\cite{eckmann13} data, respectively. 
  (c) Distributions of normalized $\bm{q}$-resolved Raman intensity $I_{\bm q}$ in BZ for D (top) and $\rm D^\prime$ (bottom) bands at $E_\mathrm{L}=\SI{2.41}{eV}$. The value of each panel is normalized by its corresponding maximum.
  The calculated defect distance is $L_\mathrm{D}=\SI{1.0}{nm}$, while the experimental value is not explicitly specified in Ref.~\cite{eckmann13}.
  (d)  Raman intensity ratio $I_\mathrm{D}/I_\mathrm{D^\prime}$ as a function of $L_\mathrm{D}$ at $E_\mathrm{L}=\SI{2.41}{eV}$. The red squares and green circles represent calculated and experimental~\cite{lucchese10,cancado11} data, respectively. Reproduced from Ref.~\citep{liu2026defectRaman} with permission. Copyright 2026 American Physical Society.}
\end{figure*}

For the $\mathrm{pd}$ ordering, the Raman tensor for phonon mode $\mu$, wavevector $\bm{q}$, and incident laser energy $E_{\mathrm{L}}$ is given by:
\begin{strip}
\begin{equation}
\overleftrightarrow{R}_{\mathrm{Uc,pd}}(\bm{q},\mu,E_{\mathrm{L}})
=\sum_{\bm{k},i,n,n',n''}
\frac{\left[D_{fn''}(\bm{k})\cdot\mathcal{M}^{\mathrm{e-d}}_{n''n'}(\bm{k}+\bm{q},-\bm{q})\right]
\left[\mathcal{M}^{\mathrm{e-ph}}_{n'n,\mu}(\bm{k},\bm{q})\cdot D^{\dagger}_{ni}(\bm{k})\right]}
{\left(E_{ni}-E_{\mathrm{L}}-\mathrm{i}\gamma\right)
\left(E_{n'i}-E_{\mathrm{L}}\pm\hbar\omega_{\mu}-\mathrm{i}\gamma\right)
\left(E_{n''f}-E_{\mathrm{L}}\pm\hbar\omega_{\mu}-\mathrm{i}\gamma\right)}.
\label{eq:uc_drr_pd}
\end{equation}
For the $\mathrm{dp}$ ordering, the corresponding tensor is
\begin{equation}
\overleftrightarrow{R}_{\mathrm{Uc,dp}}(\bm{q},\mu,E_{\mathrm{L}})
=\sum_{\bm{k},i,n,n',n''}
\frac{\left[D_{fn''}(\bm{k})\cdot\mathcal{M}^{\mathrm{e-ph}}_{n''n',\mu}(\bm{k}+\bm{q},-\bm{q})\right]
\left[\mathcal{M}^{\mathrm{e-d}}_{n'n}(\bm{k},\bm{q})\cdot D^{\dagger}_{ni}(\bm{k})\right]}
{\left(E_{ni}-E_{\mathrm{L}}-\mathrm{i}\gamma\right)
\left(E_{n'i}-E_{\mathrm{L}}-\mathrm{i}\gamma\right)
\left(E_{n''f}-E_{\mathrm{L}}\pm\hbar\omega_{\mu}-\mathrm{i}\gamma\right)}.
\label{eq:uc_drr_dp}
\end{equation}
\end{strip}
Here, $D_{ni}$ denotes the electron-photon dipole matrix element, $\mathcal{M}^{\mathrm{e-ph}}_{n'n,\mu}$ is the e-ph coupling matrix element, and $\gamma$ is the electronic energy broadening. The intermediate-state energies are denoted by $E_{ni}$, $E_{n'i}$, and $E_{n''f}$. For the closed Raman cycle, the electron recombines with the hole, and we therefore set the final electronic state $f$ equal to the initial state $i$. The $\pm$ sign distinguishes Stokes and anti-Stokes processes.

The total Raman tensor within the selected carrier pathway is the coherent sum of the two time orderings,
\begin{equation}
\overleftrightarrow{R}_{\mathrm{Uc}}
=
\overleftrightarrow{R}_{\mathrm{Uc,pd}}
+
\overleftrightarrow{R}_{\mathrm{Uc,dp}}.
\label{eq:uc_drr_total}
\end{equation}
The Raman intensity from this $ee$ pathway is obtained by contracting the tensor with the incident and scattered light polarization vectors and summing over phonon wavevectors and branches:
\begin{equation}
I_{\mathrm{Uc}}^{(ee)}(\omega_{\mathrm{RS}})
\propto
\sum_{\bm{q},\mu}
\left|
\bm{P}_{\mathrm{s}}^{\dagger}\cdot
\overleftrightarrow{R}_{\mathrm{Uc}}(\bm{q},\mu,E_{\mathrm{L}})\cdot
\bm{P}_{\mathrm{i}}
\right|^2
L\!\left(\hbar\omega_{\mathrm{RS}}\pm\hbar\omega_{\mu}\right),
\label{eq:uc_drr_intensity}
\end{equation}
where $\bm{P}_{\mathrm{i}}$ and $\bm{P}_{\mathrm{s}}$ are the incident and scattered polarization vectors, respectively, and $L$ is a Lorentzian function representing the spectral broadening. The two time-ordering amplitudes are added coherently before taking the absolute square, so they can interfere.

Recent first-principles calculations apply the Uc-DRR approach to monolayer graphene and MoSe$_2$, alongside a complementary calculation using defect-containing supercells~\citep{liu2026defectRaman}. In the Uc-DRR approach, the electronic states and phonons of the pristine crystal are retained, and the defect enters through the calculated elastic scattering matrix element. This approach describes defect-activated finite-$\bm{q}$ phonons and reproduces the characteristic D and D$'$ bands in graphene. However, for monolayer MoSe$_2$, the Uc-DRR approach captures only some defect-induced Raman modes. As shown in Fig.~\ref{fig:mose2}(a), the defect-induced Raman peaks, including $D_1$, $D_4$, and $D_5$ are obtained by the Uc-DRR approach, while $D_2$ and $D_3$ are not. This is because $D_2$ and $D_3$ occur in the phonon gap region (see Fig.~\ref{fig:mose2}(b)), which thus cannot be obtained by the Uc-DRR approach. To capture the $D_2$ and $D_3$ peaks, we use the Sc-SRR approach. The Sc-SRR approach explicitly includes the modified structure and vibrational modes around the defect. It can therefore describe localized phonon modes within a phonon-band gap and defect-induced frequency shifts of first-order Raman modes, which are absent from the pristine phonon dispersion. The two approaches describe distinct contributions to the spectrum and are most useful when applied together~\citep{liu2026defectRaman}, as shown in Fig.~\ref{fig:mose2}(a).

\begin{figure*}[ht]
  \centering
  \includegraphics[width=0.85\textwidth]{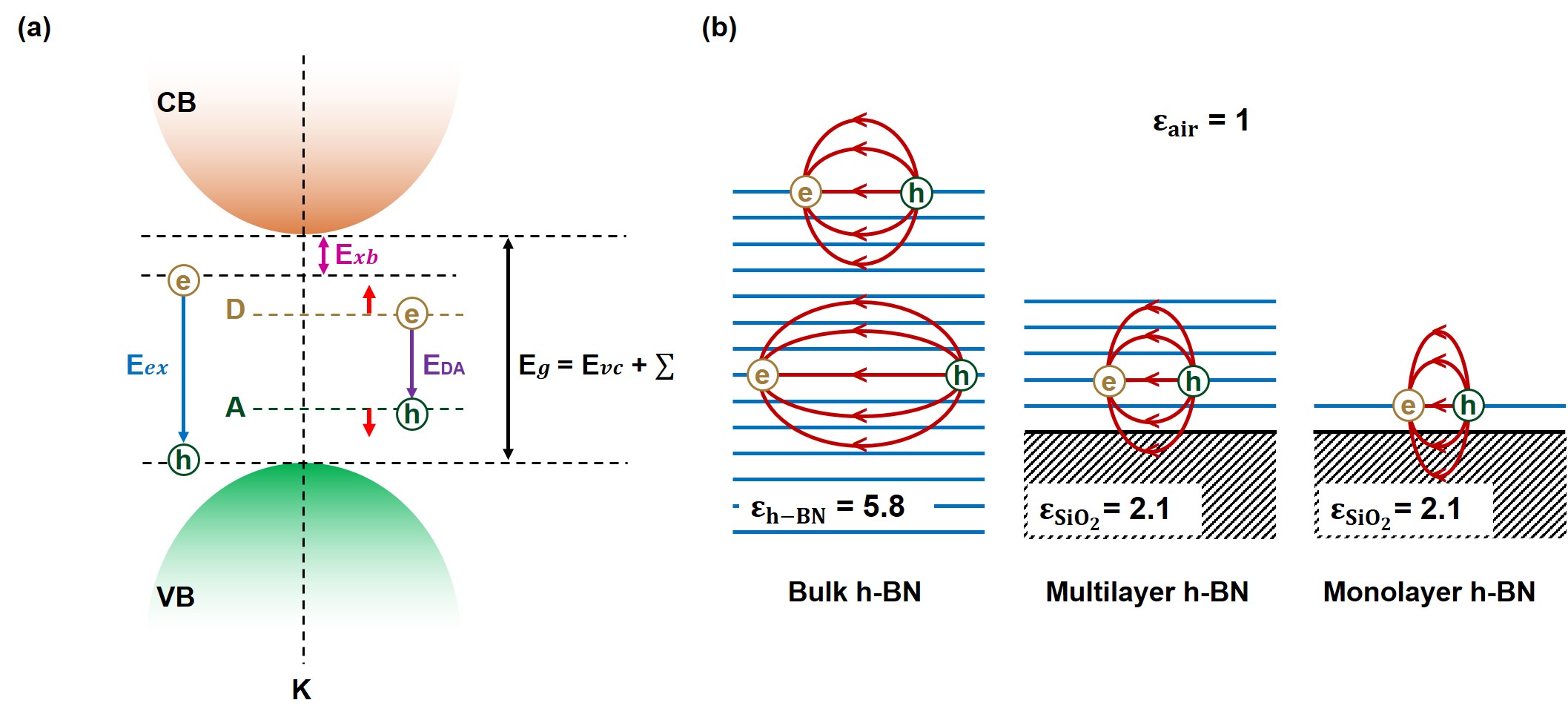}
  \caption{Environmental effects on excitonic energies in hBN.
  (a) Schematic energy diagram distinguishing the band-edge exciton
  energy $E_{\mathrm{ex}}$ from the defect-induced donor-acceptor
  exciton energy $E_{\mathrm{DA}}$. The renormalized band gap and
  exciton binding energy contribute to the optical transition energies.
  (b) Real-space illustrations of electron-hole interactions in bulk,
  multilayer, and monolayer hBN, showing the influence of layer
  position and the surrounding dielectric environment. Reproduced from
  Ref.~\citep{thanh2025hbn} with permission. Copyright 2025 American
  Physical Society.}
  \label{fig:hbn_dap_environment}
\end{figure*}

This distinction also matters when assigning experimentally observed peaks. A feature activated by defect scattering of a pristine finite-wavevector phonon can overlap spectrally with a localized mode or a pristine second-order Raman band. Their laser-energy dependence, polarization response, and calculated atomic displacement patterns can provide complementary evidence for their origins. For example, helicity-resolved and linearly polarized Raman calculations help distinguish modes with different symmetries, even when their Raman shifts are similar~\citep{liu2026defectRaman}.

Calculating the elastic scattering matrix element directly thus advances defect-induced Raman theory from assigning peak positions to predicting the microscopic contributions to spectral intensity. The current calculations demonstrate this framework for charge-neutral defects. Extending it to charged defects in 2D materials remains challenging because their long-range electrostatic potentials require larger supercells and careful treatment of dielectric screening. Combining accurate defect scattering with defect-modified phonons, dielectric-environment effects, and quantitative comparison with experiment will be important for using Raman spectra to identify and characterize point defects in 2D materials.

\subsection{Photoluminescence spectra from 2D donor-acceptor states}

Photoluminescence (PL) from a donor-acceptor (DA) pair occurs when an
electron localized at a donor recombines radiatively with a hole
localized at an acceptor. In a simplified pair model, the photon energy
depends on the donor-acceptor separation $R$~\cite{thanh2025hbn}:
\begin{equation}
E_{\mathrm{PL}}(R)
\simeq E_{\mathrm{g}}-E_{\mathrm{D}}-E_{\mathrm{A}}
+W_{\mathrm{DA}}(R),
\end{equation}
where $E_{\mathrm{g}}$ is the host band gap, $E_{\mathrm{D}}$ and
$E_{\mathrm{A}}$ are the donor and acceptor binding energies, and
$W_{\mathrm{DA}}(R)$ is the magnitude of the attractive Coulomb
interaction between the electron and hole. In a bulk semiconductor,
this interaction is often approximated by a term proportional to
$1/(\varepsilon R)$. In a 2D material, however, screening is nonlocal
and depends on the dielectric environment, so the pair interaction
cannot in general be described by a single dielectric constant. The
pair separation and configuration also affect the wavefunction overlap
and hence the transition intensity. An ensemble of pairs with different
separations, defect types, and local environments may therefore produce
a broad PL band or multiple emission lines.

Figure~\ref{fig:hbn_dap_environment} illustrates how the dielectric
environment can influence both band-edge excitons and defect-induced
DA emission in hBN~\cite{thanh2025hbn}. Panel~(a) distinguishes the band-edge exciton
energy $E_{\mathrm{ex}}$ from the defect-induced exciton energy
$E_{\mathrm{DA}}$, associated with recombination between donor and
acceptor states. It also depicts the contributions of the renormalized
band gap and exciton binding energy. Panel~(b) compares the
electron-hole interaction in bulk, multilayer, and monolayer hBN,
where the surrounding dielectric media and the position of a layer
within the stack alter the electrostatic environment. This schematic
provides a useful framework for interpreting thickness-dependent
optical transitions, although it does not represent the full
nonlocal defect potential.

Hexagonal boron nitride (hBN) provides an example of how these effects
appear in experiment. Donor-acceptor recombination has been associated
with defect-induced PL in the ultraviolet and deep-ultraviolet spectral
ranges. In bulk hBN, the observed emission contains several subpeaks
whose energy separations are comparable to the optical-phonon energy;
these features have been interpreted as phonon replicas of DA
recombination. In thinner hBN films, the subpeaks are not resolved and
the PL spectra are broader. Measurements across monolayer, multilayer,
and bulk samples show that the defect-induced PL peak positions shift
to higher energy with increasing film thickness. The concurrent
thickness dependence of the band-to-band exciton absorption indicates
that the optical response is also sensitive to the dielectric
environment. A correlation between the DA emission energy
$E_{\mathrm{DA}}$ and the band-edge exciton energy $E_{\mathrm{ex}}$
was observed; the variation in $E_{\mathrm{DA}}$ is larger, consistent
with the stronger environmental sensitivity of the localized
defect-induced exciton~\citep{thanh2025hbn}. These results show that
the PL energy of a defect pair need not be determined by its local
chemical identity alone, but can also depend on the surrounding layers
and dielectric environment.

Carbon incorporation in monolayer boron-carbon-nitrogen (BCN)
provides a complementary example in which the local defect
configuration can modify donor- and acceptor-like states and thereby
contribute to multiple optical features~\cite{tseng2026wafer}. In the measured BCN sample,
broad PL components are observed near 2.16, 3.47, and 4.40~eV, while
corresponding absorption features occur near 2.30, 3.65, and 4.56~eV.
The resulting Stokes shifts of approximately 0.14-0.18~eV are
comparable to the optical-phonon energy of hBN. This comparison is
consistent with phonon-assisted relaxation, although it does not by
itself establish a unique relaxation pathway for each feature.

\begin{figure*}[t]
  \centering
  \includegraphics[width=0.85\linewidth]{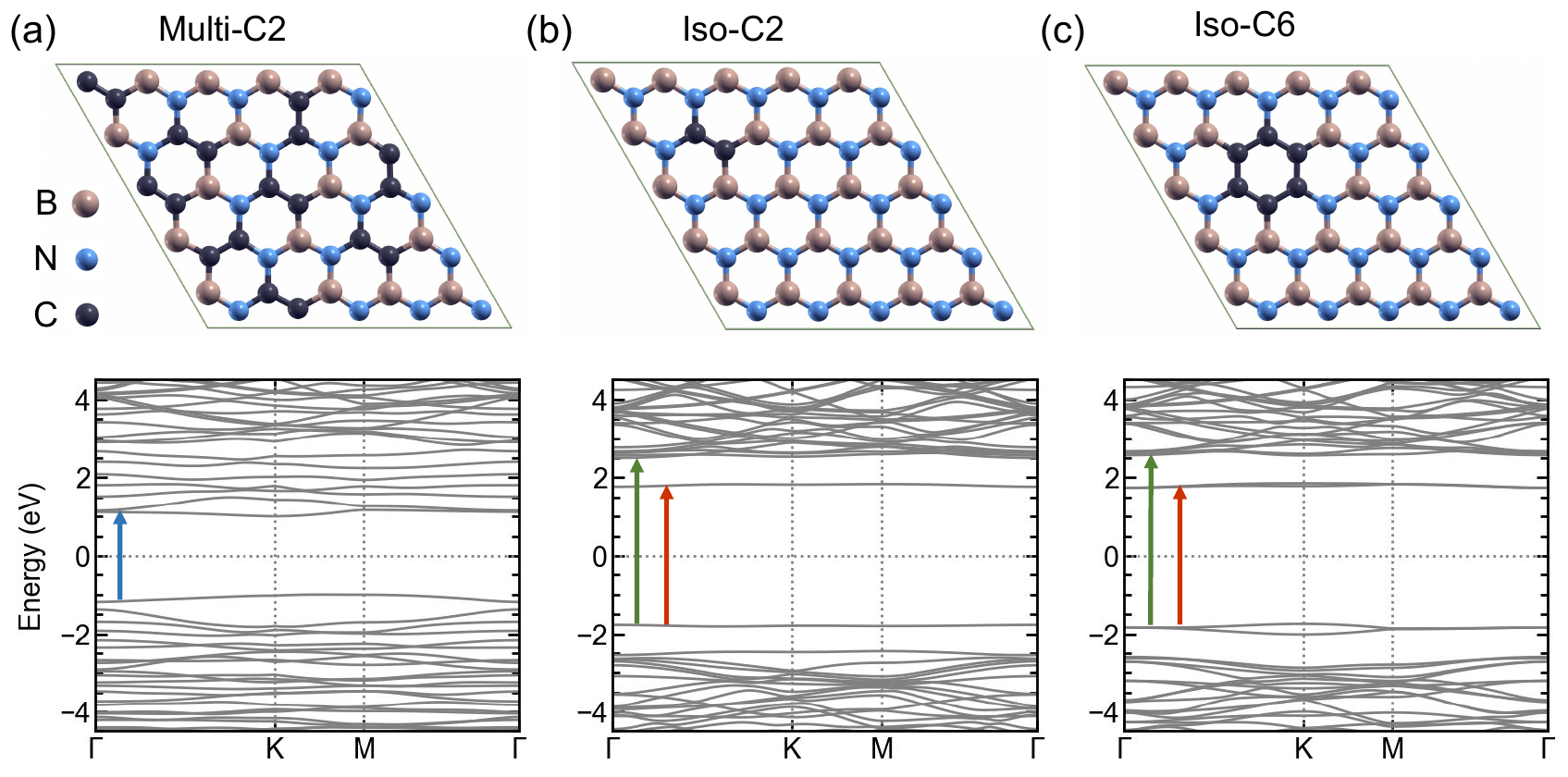}
  \caption{Atomic structures (top) and calculated electronic band
  structures (bottom) of three representative monolayer BCN
  configurations in a $5\times5$ hBN supercell: (a) nine carbon dimers
  (Multi-C2), (b) one isolated carbon dimer (Iso-C2), and (c) one
  isolated six-membered carbon ring (Iso-C6). The blue arrow in (a),
  and the red and green arrows in (b) and (c), respectively, indicate
  candidate transitions associated with the experimentally observed
  absorption features at 2.30, 3.65, and 4.56~eV. The Fermi energy is
  set to $E=0$~eV in each panel. Adapted from Ref.~\citep{thanh2026bcn}.}
  \label{fig:BCN_bands}
\end{figure*}

To explore how the carbon configuration affects the electronic states,
first-principles calculations were performed for representative carbon
defects in a $5\times5$ hBN supercell. Figure~\ref{fig:BCN_bands}
compares three configurations: a periodic arrangement of nine carbon
dimers (Multi-C2), one isolated carbon dimer (Iso-C2), and one isolated
six-membered carbon ring (Iso-C6). The selected structures are
semiconducting, with the Fermi energy set to zero in each panel.
Carbon substitution at B and N sites gives rise to donor-like and
acceptor-like states, respectively, but the dispersion and energy
separation of these states depend on the local configuration.

In Iso-C2, the donor- and acceptor-like states are nearly
dispersionless, consistent with states localized around the substituted
carbon atoms. In Iso-C6, the corresponding states show weak dispersion,
which may reflect electronic coupling within the carbon ring. In
Multi-C2, interactions among states associated with neighboring dimers
produce multiple weakly dispersive bands and reduce the separation
between the highest acceptor-like state and the lowest donor-like state
relative to the isolated-defect configurations. Thus, even when the
same broad classes of donor- and acceptor-like states are present,
their energies and degree of dispersion are sensitive to the defect
arrangement.

The colored arrows in Fig.~\ref{fig:BCN_bands} mark candidate
transitions whose energy differences are close to the measured
absorption features. The Multi-C2 model provides a candidate transition
near 2.30~eV, whereas the Iso-C2 and Iso-C6 models provide candidate
transitions near 3.65 and 4.56~eV, respectively. This correspondence
suggests that carbon dimers and rings may contribute to the optical
response of BCN. It should be interpreted as a set of candidate
assignments rather than a unique identification: the band structures
indicate possible transition energies, but do not alone establish the
microscopic origin of each measured PL component or its relaxation
pathway.

The calculations represent selected local configurations rather than
the full distribution of carbon defects in the sample. The measured
spectrum may therefore contain contributions from multiple structures,
including configurations not included in the present supercells.
Moreover, deviations from a 1:1 B:N ratio and additional defects such
as vacancies may shift the Fermi level toward donor- or acceptor-like
states, or otherwise alter the electronic structure. Together, the
hBN and BCN examples show that dielectric-environment effects and
defect-configuration-dependent electronic structure can both influence
optical spectra in 2D materials. Quantitative interpretation of the
PL additionally requires treatment of excited-state relaxation and
vibronic coupling.



\section{Outlook}
In this section, we highlight several emerging directions for advancing the understanding and prediction of electron-defect interactions, including charged-defect scattering, electron-phonon-defect coupling, exciton-defect coupling, and AI-driven approaches.

\subsection{Charged-defect scattering for 2D materials}
Charged defects are particularly challenging scattering centers in 2D materials because their long-range Coulomb potentials decay much more slowly than the short-range perturbations associated with neutral defects. Direct evaluation of such potentials using conventional DFT supercells therefore requires prohibitively large simulation cells. Moreover, dielectric screening in 2D materials is strongly nonlocal and anisotropic and can depend sensitively on carrier concentration, temperature, and the out-of-plane charge distribution. Conventional approaches often approximate the bare defect charge as a point charge and employ simplified screening models, such as Thomas-Fermi screening (see Sec. 3.2) or an analytic model (see Sec. 5.2), which may introduce substantial uncertainties in the defect potential $\Delta V_{\mathrm{def}}$ and consequently in the calculated scattering rates and transport properties.

Recent progress has substantially improved this situation. Guo \textit{et al.}~\cite{guo2025accurate} developed a first-principles method in which the long-range part of $\Delta V_{\mathrm{def}}$, e.g., the change of the Hartree potential, $\Delta V_H$, between the neutral defective structure and the charged structure, is given by
\begin{equation}
\Delta V_H=W\Delta \rho_b
\end{equation}
where $\rho_b$ is the bare (unscreened) excess charge density introduced by the charged defect and $W$ is the screened Coulomb interaction, which is defined by~\cite{guo2025accurate}
\begin{equation}
W=
\left[I-v\left(\chi_b^0+\chi_c^0\right)\right]^{-1}v,
\end{equation}
where $I$ is the identity operator, $v$ is the bare Coulomb interaction, and $\chi_b^0$ and $\chi_c^0$ describe the polarization responses of bound electrons and free carriers, respectively. This separation between the localized bare defect charge $\Delta\rho_b$ and the long-range screened potential $W$ is particularly powerful since $\Delta\rho_b$ needs to be determined only within a relatively small defect supercell, whereas $W$ can subsequently be reconstructed for arbitrarily large length scales, ultimately approaching the isolated-defect limit.

\begin{figure}[t]
    \centering
    \includegraphics[width=1.0\linewidth]{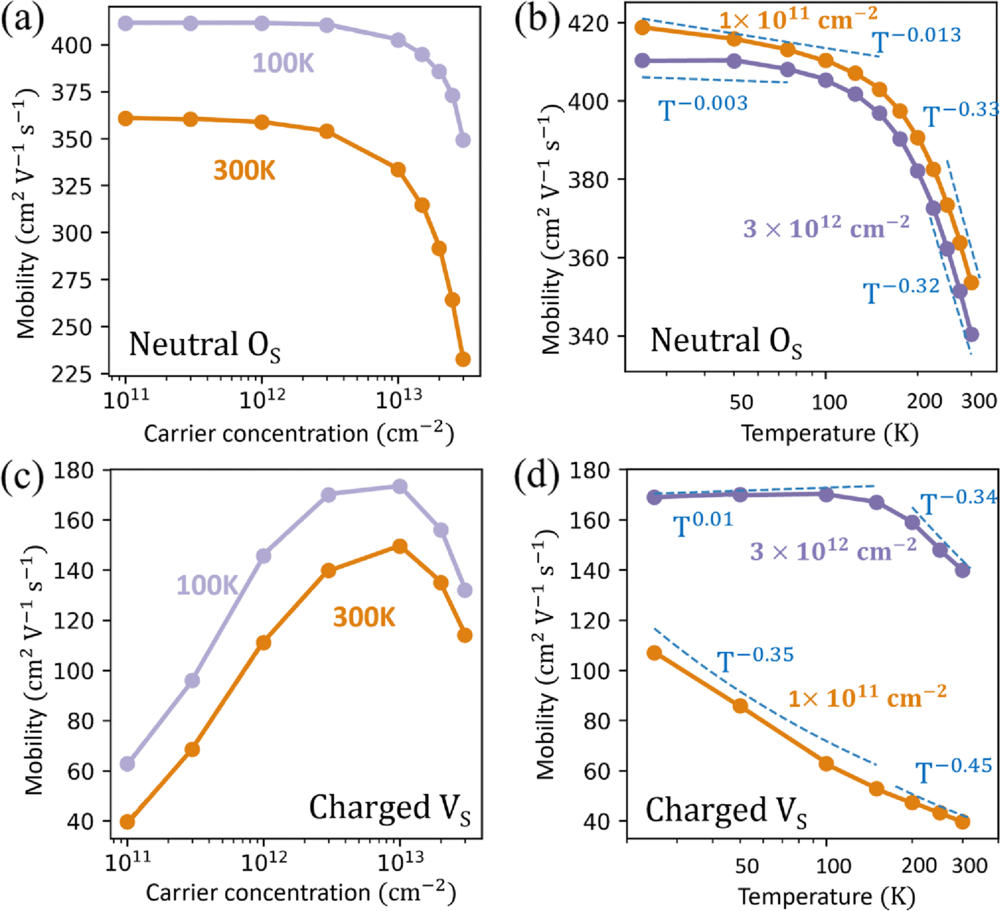}
    \caption{Electron mobility of monolayer MoS$_2$ as a function of carrier concentration and temperature, under scattering by charge-neutral O substitutes of S (O$_\text{S}$) (a), (b), and -1 charged S vacancies (V$_\text{S}$) (c), (d). The temperature-dependent fitting is shown as blue dashed lines. Reproduced from Ref.~\citep{guo2025accurate} with permission. Copyright 2025 American Physical Society.}
    \label{fig:charged}
\end{figure}

For 2D materials, an important advance is the explicit treatment of the finite out-of-plane carrier distribution and quasi-2D screening rather than assuming an infinitely thin homogeneous electron gas. The method was demonstrated for negatively charged sulfur vacancies, $V_{\mathrm S}^{-1}$, in monolayer MoS$_2$, where a bare charge obtained from a $6\times6$ supercell successfully reproduced the long-range potential of a larger $9\times9$ calculation~\cite{guo2025accurate}. The resulting mobility calculations further showed that charged-defect scattering can exhibit pronounced carrier-density and temperature dependence due to free-carrier screening, in contrast to the much weaker temperature dependence typically associated with neutral defects (see Fig.~\ref{fig:charged}). 

Despite these advances, several important challenges remain. Current formulations generally assume that the defect has a negligible effect on the host dielectric response, allowing $\chi_b^0$ and $\chi_c^0$ to be evaluated from the pristine crystal. Extending the method to strongly perturbing defects, multiband systems, multilayers, heterostructures, and moir\'e materials will require a more general description of the spatially dependent polarizability. In realistic devices, screening from substrates, encapsulation layers, metallic gates, and nearby conducting channels should also be incorporated, since these environments can substantially modify the long-range defect potential.

A further challenge is to move beyond static defects with fixed charge states. In realistic conditions, defect occupations can change through carrier capture and emission and may depend on the Fermi level, temperature, illumination, and nonequilibrium carrier populations. Developing a unified first-principles framework that combines defect charge-state thermodynamics, carrier capture and emission, dynamical screening, and defect-limited transport therefore remains an important direction for future studies of 2D materials.

\subsection{Electron-phonon-defect coupling}
Electron-phonon (e-ph) and electron-defect (e-d) interactions are commonly calculated as independent scattering mechanisms. Within this approximation, their contributions to carrier transport are often combined by using Matthiessen’s rule as:
\begin{equation}
\frac{1}{\tau_{\mathrm{tot}}}
\simeq
\frac{1}{\tau_{\mathrm{e-ph}}}
+
\frac{1}{\tau_{\mathrm{e-d}}}.
\end{equation}
However, a point defect not only scatters electrons but also modifies the local lattice dynamics, including phonon frequencies, eigenvectors, linewidths, and potentially localized vibrational modes. At the same time, the defect perturbs and broadens the electronic states. Consequently, the e-ph interaction in a defective material is not necessarily identical to that of the pristine crystal, and treating e-ph and e-d scattering as completely independent processes may neglect their coupled contribution.

Recent work has begun to address this problem using a many-body perturbation theory (MBPT) framework for coupled electron-phonon-defect interactions \cite{giustino2017electron,marini2015many}. The central idea is to extend the conventional electron-phonon self-energy by replacing the pristine electron and phonon quasiparticles with their defect-modified spectral functions. Schematically, the resulting self-energy can be viewed as
\begin{equation}
\Sigma_{\mathrm{e-ph}}^{\mathrm{def}}
\sim
|g|^2
G^{\mathrm{def}}
D^{\mathrm{def}},
\end{equation}
where $G^{\mathrm{def}}$ and $D^{\mathrm{def}}$ denote the defect-modified electronic and phononic Green's functions, respectively, and $g$ represents the e-ph coupling. Equivalently, the corresponding electronic and phononic spectral functions contain the broadening and renormalization introduced by defects. This formulation therefore allows the influence of defects on e-ph scattering to be incorporated directly rather than treating the two interactions as independent perturbations.

Once the defect-modified spectral functions and self-energies are obtained, the electrical conductivity can be evaluated within the Kubo-Greenwood formalism. Such an approach goes beyond a simple application of Matthiessen's rule and provides a route toward describing transport when e-ph and e-d interactions become intertwined.

Several challenges nevertheless remain before such calculations can become broadly applicable. Point defects break translational symmetry and generally require large supercells, while accurate electron-phonon calculations require dense sampling of both electronic and phonon momentum spaces. Combining the two therefore introduces a substantial computational burden. Efficient interpolation or reduced-order representations of defect-modified electron-phonon interactions will be essential for realistic materials and dilute defect concentrations. In addition, configurational averaging over different defect positions, concentrations, and defect types will be required to connect calculations with experimentally disordered materials.

More fundamentally, strongly perturbing defects can generate localized electronic states and local vibrational modes for which the conventional quasiparticle description may become inadequate. Extending coupled electron-phonon-defect theories to include strong coupling, vertex corrections, dynamical screening, and nonequilibrium carrier and phonon populations therefore represents an important future direction. These effects may be especially significant in 2D materials, where reduced dimensionality, strong electron-phonon interactions, and enhanced sensitivity to individual defects can make the interplay among electrons, phonons, and defects particularly pronounced.

\subsection{Exciton-defect coupling}
Exciton-defect interactions are particularly important in 2D semiconductors, where reduced dielectric screening gives rise to strongly bound excitons that dominate the low-energy optical response. Atomic defects can perturb both the electron and hole components of an exciton, leading to exciton scattering, localization, hybridization with defect states, and the formation of defect-bound excitons. These effects can strongly modify optical absorption, photoluminescence, valley selectivity, and exciton lifetimes, and may also provide localized states for quantum-light emission.

First-principles GW-Bethe-Salpeter equation (GW-BSE) calculations have demonstrated that point defects can substantially modify the excitonic structure of 2D materials. For example, chalcogen vacancies in monolayer transition-metal dichalcogenides can introduce strongly bound defect excitons and hybridize with pristine excitonic states, thereby modifying the valley-selective optical response~\cite{refaely2018defect}. However, conventional GW-BSE calculations for defective supercells are computationally demanding and are generally restricted to relatively small supercells. Consequently, the simulated defect concentrations are often much larger than those encountered experimentally, and interactions between periodically repeated defects may obscure the isolated-defect limit.

The exciton-defect scattering matrix element is defined by:
\begin{equation}
\begin{split}
\mathcal{M}_{S\bm{Q}+\bm{q},S'\bm{Q}}^{\text{exciton}}=
\sum_{\bm k}
\Bigg[
&\sum_{vcc'}
A^{S*}_{c\bm k+\bm Q+\bm q,v\bm k}
A^{S'}_{c'\bm k+\bm Q,v\bm k}
\mathcal{M}_{cc'}^{\text{def}}(\bm k+\bm Q,\bm q)
\\
-&
\sum_{cvv'}
A^{S*}_{c\bm k+\bm Q,v\bm k-\bm q}
A^{S'}_{c\bm k+\bm Q,v'\bm k}
\mathcal{M}_{v'v}^{\text{def}}(\bm k-\bm q,\bm q)
\Bigg],
\label{eq:exciton_defect}
\end{split}
\end{equation}
where $\bm Q$ and $\bm Q+\bm q$ are the initial and final exciton center-of-mass crystal momenta, $\bm q$ is the momentum transferred by the defect, $S'$ and $S$ are the initial exciton band indices, $A^{S}_{c\bm k+\bm Q,v\bm k}$ is the exciton envelope function obtained by solving the BSE, with $c$ and $v$ denoting conduction- and valence-band states, respectively. The quantity $\mathcal{M}_{nm}^{\text{def}}(\bm k,\bm q)$ is the e-d matrix element between the single-particle states $|m,\bm k\rangle$ and $|n,\bm k+\bm q\rangle$. The first term in Eq.~\eqref{eq:exciton_defect} describes scattering of the electron component of the exciton by the defect, whereas the second term describes scattering of its hole component. The opposite sign of the hole contribution reflects the fact that a hole corresponds to the absence of a valence electron.

\begin{figure}[t]
    \centering
    \includegraphics[width=1.0\linewidth]{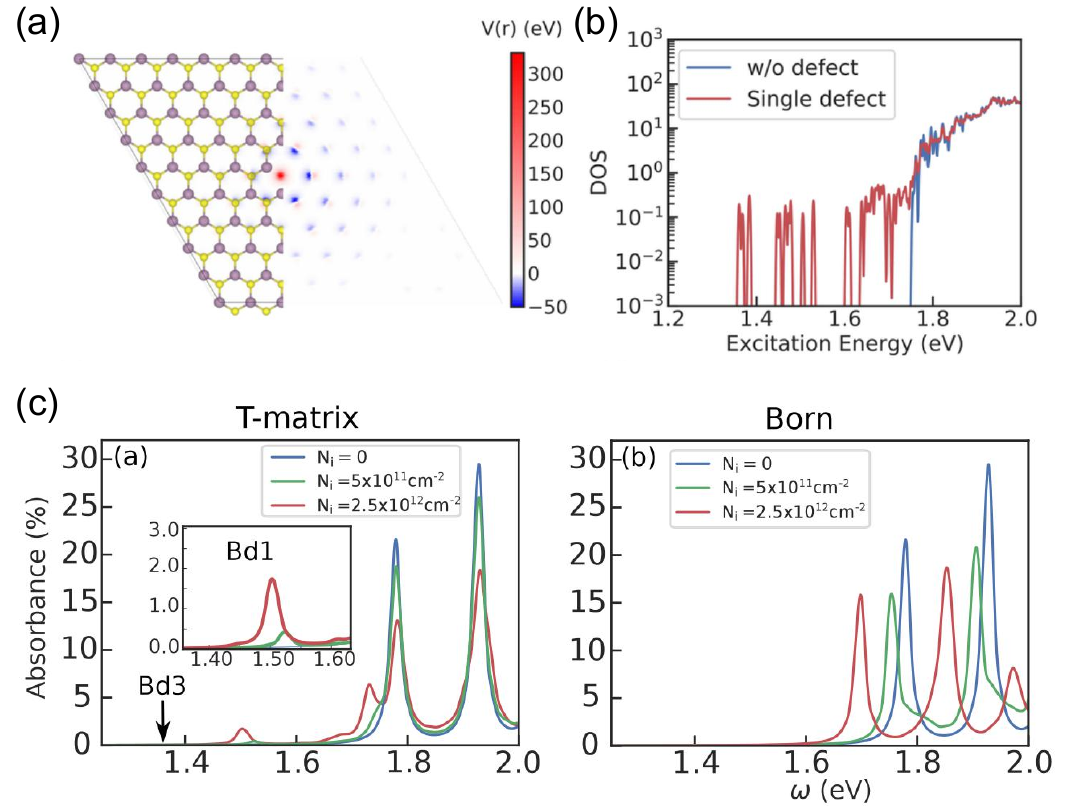}
    \caption{(a) Intensity map of a single S-vacancy defect potential $V(\bm r)$ in a $9 \times 9$ supercell overlaid with the atomic structure, in which purple balls are Mo atoms and yellow balls are S atoms. (b) Density of exciton states without (blue) and with (red) the single S-vacancy defect in the supercell. (c) Calculated optical absorbance for monolayer MoS$_2$ within the T-matrix (right) and Born (left) approximations at several defect densities $N_i$ at 300 K. The inset shows a zoom-in view near 1.5 eV for the Bd1 defect-bound exciton, while the arrow indicates the energy of the Bd3 defect-bound exciton. Reproduced from Ref.~\citep{chan2026exciton} with permission. CC BY 4.0.}
    \label{fig:exciton}
\end{figure}

The exciton-defect matrix element in Eq.~\eqref{eq:exciton_defect} can be calculated from the e-d matrix elements obtained from DFT together with the exciton wave functions obtained from the BSE. This matrix element describes the coupling between pristine exciton states induced by a defect. However, when the interaction is treated only within the Born approximation (i.e., the defect potential is treated as a weak perturbation), it mainly describes exciton scattering, energy shifts, and broadening, and is generally insufficient to capture the formation of a defect-bound exciton. To describe such bound states, the exciton-defect matrix element can instead be used as the interaction $V$ in the T-matrix formalism~\cite{kaasbjerg2020atomistic},
\begin{equation}
T(\omega)=V+VG_0(\omega)T(\omega),
\end{equation}
where $\omega$ is the excitation frequency and $G_0(\omega)$ is the Green's function of the pristine exciton. Then, the T-matrix self-energy is given by $\Sigma^T(\omega)=N_iT(\omega)$, where $N_i$ is the number of defects~\cite{kaasbjerg2020atomistic}.

The T-matrix effectively includes repeated scattering of an exciton from the same defect, \(V+VG_0V+VG_0VG_0V+\cdots\)~\cite{bruus2004many}, which allows an exciton to become localized around the defect and form a defect-bound exciton. For sulfur vacancies in monolayer MoS\(_2\), this approach yields additional defect-bound exciton features in the calculated absorption that cannot be reproduced by the Born approximation~\citep{chan2026exciton}, as shown in Fig.~\ref{fig:exciton}.

Despite these advances, several challenges remain. The present first-principles treatment has primarily been demonstrated for neutral point defects and relatively simple monolayer systems. Extending it to charged defects will require a consistent treatment of the long-range screened defect potential together with the electron-hole interaction, particularly in 2D materials where screening depends strongly on the surrounding dielectric environment. Defect charge states, substrates, encapsulation, strain, and local electrostatic fields may all modify both the excitonic wave function and the defect potential.

Another important direction is the inclusion of dynamical processes. Real excitons can be captured by or released from defects through phonon-assisted transitions, while defects can mediate scattering between bright, dark, intervalley, and spin-forbidden excitons. A unified treatment of exciton-defect and exciton-phonon interactions will therefore be required to predict defect-assisted relaxation, trapping, thermalization, and recombination rates. Such developments are especially relevant for understanding defect-related photoluminescence and single-photon emission, for which recent experiments are beginning to correlate individual atomic defects directly with spatially localized excitonic emission.

More broadly, extending first-principles exciton-defect calculations to realistic defect densities, multiple defect species and charge states, heterogeneous dielectric environments, and nonequilibrium exciton populations remains computationally demanding. Efficient interpolation of exciton-defect matrix elements, disorder averaging, and reduced-order or machine-learning approaches may therefore play an important role in connecting atomic-scale defect physics with experimentally measurable optical spectra and exciton dynamics.

\subsection{AI-driven electron-defect coupling}
First-principles calculations of e-d interactions remain computationally demanding because defect calculations typically require large supercells, structural relaxation, and dense sampling of electronic states. The computational cost increases rapidly when many defect species, charge states, concentrations, and atomic configurations need to be considered. Artificial intelligence and machine-learning (ML) methods therefore offer a promising route to accelerate e-d calculations, particularly by learning the electronic Hamiltonian or defect-induced perturbation from a limited set of first-principles calculations~\cite{fu2025ai,kiyohara2025machine,kumagai2026machine}.

Recent developments in machine-learned electronic Hamiltonians, such as DeepH~\cite{li2022deep}, DeepH-E3~\cite{gong2023general}, and HamGNN~\cite{zhong2023transferable}, have shown that symmetry-preserving neural networks can reproduce DFT Hamiltonian matrices with high accuracy and transfer to electronic-structure calculations of large, structurally complex systems. Similar approaches have also been applied to defective systems, including the construction of machine-learned tight-binding Hamiltonians for defects and scalable electronic-structure calculations of large defect-containing supercells~\cite{schattauer2022machine}. These developments suggest a natural extension toward AI-driven electron-defect coupling.

By assuming the kinetic energies of the defective-supercell and a pristine-supercell are the same, we have $\Delta V_{\mathrm{def}} =\Delta H_{\mathrm{def}}$. Then, the electron-defect coupling in Eq.~\eqref{eq:2} can be rewritten in terms of the Hamiltonian as follows: 
\begin{equation}
\mathcal{M}_{mn}^{\mathrm{def, ML}}(\bm{k},\bm{k}')=
\left\langle
m\bm{k}'
\left|
\Delta H_{\mathrm{def}}^{\mathrm{ML}}
\right|
n\bm{k}
\right\rangle ,
\label{eq:ML_ed}
\end{equation}
where $\Delta H_{\mathrm{def}}^{\mathrm{ML}} = H_{\mathrm{def}}^{\mathrm{ML}}-H_{\mathrm{pris}}^{\mathrm{ML}}$. Here, $H_{\mathrm{def}}^{\mathrm{ML}}$ and $H_{\mathrm{pris}}^{\mathrm{ML}}$ are the Hamiltonians of the defective-supercell and pristine-supercell, which can be obtained using the ML models, such as DeepH-E3 and HamGNN. In this way, the ML model replaces the computationally expensive construction of the electronic Hamiltonian, while the subsequent evaluation of scattering rates and transport coefficients can retain the conventional quantum-mechanical framework. This strategy is conceptually similar to recent ML approaches for electron-phonon coupling, in which equivariant neural networks are used to predict electronic Hamiltonians and their derivatives while the interaction matrix elements and transport properties are subsequently evaluated using established many-body and transport theories~\cite {zhong2024accelerating}.

Several challenges must nevertheless be addressed before AI-driven electron-defect coupling calculations become predictive. The e-d matrix elements depend not only on the local atomic geometry but also on defect type, charge state, electronic band character, spin-orbit coupling, and, for charged defects, the dielectric and carrier-screening environment. Furthermore, unlike scalar quantities such as formation energies, Hamiltonians and scattering matrix elements must obey rotational, translational, Hermitian, and gauge-related constraints. Physics-informed and equivariant architectures are therefore particularly important for maintaining these symmetries.

Another major limitation is the availability of high-quality training data. The space of possible defects is extremely large, including vacancies, substitutions, interstitials, antisites, defect complexes, different charge states, and locally reconstructed configurations. Active learning and uncertainty quantification may therefore be essential for identifying the most informative first-principles calculations and systematically expanding the training set. In the longer term, transferable or foundation electronic-structure models capable of describing both pristine and defective environments could enable electron--defect matrix elements to be evaluated for thousands of defect configurations at a small fraction of the cost of direct DFT calculations.

AI-driven electron--defect coupling is still early compared with machine-learning approaches for energies, forces, and electronic structures. Nevertheless, recent progress in transferable machine-learned Hamiltonians and interaction matrix elements provides a promising foundation. Combining such models with Wannier interpolation, first-principles defect potentials, and Boltzmann or Green's-function transport methods could ultimately enable high-throughput predictions of defect-limited carrier transport across complex materials and realistic defect populations.

\section{Summary}
Point defects play a dual role in two-dimensional materials: they can limit carrier transport and optical performance through scattering and localization, while also providing opportunities to deliberately engineer electronic, vibrational, and optical properties. In this Review, we have connected the atomic origin of point defects with the defect-induced potential, electron–defect coupling, and ultimately measurable functional properties. We place particular emphasis on first-principles approaches for evaluating electron–defect matrix elements and their applications to carrier mobility, defect tolerance, electron cloaking, thermoelectric transport, defect-induced Raman scattering, and defect-related optical transitions.

Despite substantial progress, a predictive description of defects in realistic 2D materials remains challenging, particularly for charged defects, complex dielectric environments, and situations where electron–defect, electron–phonon, and exciton–defect interactions become strongly coupled. Continued advances in first-principles methods, efficient interpolation techniques, many-body theory, and machine-learning models should make it possible to treat increasingly realistic defect populations and larger material spaces. Such developments may ultimately shift the role of point defects from unavoidable imperfections toward controllable atomic-scale elements for designing the electronic, transport, and optical functionalities of two-dimensional materials.

\medskip
\noindent\textbf{Acknowledgements} \par 
\noindent N.T.H. acknowledges the financial support from the Frontier Research Institute for Interdisciplinary Sciences from Tohoku University. T.Y. acknowledges the support from the National Key R\&D Program of China (2022YFA1203900), the Strategic Priority Research Program of CAS (Grant No. XDB0460000) and the International Partnership Program of Chinese Academy of Sciences (Grant No. 172GJHZ2024044MI). J.Q.H. acknowledges the support from National Natural Science Foundation of China (Grant No. 12404213). H.H.G. acknowledges the support from the Department of Science and Technology of Liaoning Province (Grant No. 2025-MSLH-444), and the Department of Education of Liaoning Province (Grant No. LJ212610148025).

\medskip

\bibliographystyle{unsrtnat}

\end{document}